\documentclass[preprint2]{proto}
\usepackage{times}
\usepackage{amsmath}
\newcommand{\refs}{\par\noindent\hangindent=1pc\hangafter=1}
\newcommand{\mj}{M$_{\text{Jup}}$}

\newcommand{\me}{M$_{\oplus}$}

\newcommand{\teq}{$T_{\rm eq}$}
\def\ltsima{$\; \buildrel < \over \sim \;$}
\def\simlt{\lower.5ex\hbox{\ltsima}}
\def\gtsima{$\; \buildrel > \over \sim \;$}
\def\simgt{\lower.5ex\hbox{\gtsima}}

\def\msun{{\,{\rm M}_\odot}}

\newcommand\mearth{{\,{\rm M}_{\oplus}}}

\def\del#1{{}}
\usepackage[usenames,dvipsnames]{xcolor}

\usepackage[normalem]{ulem}

\begin{document}

\title{\textbf{\LARGE Giant Planet Formation, Evolution, and Internal Structure}}

\author {\textbf{\large Ravit Helled}}
\affil{\small\em Tel-Aviv University}
\author {\textbf{\large Peter Bodenheimer}}
\affil{\small\em University of California, Santa Cruz}
\author {\textbf{\large Morris Podolak}}
\affil{\small\em Tel-Aviv University}
\author {\textbf{\large Aaron Boley }}
\affil{\small\em University of Florida \& The University of British Columbia}
\author {\textbf{\large Farzana Meru}}
\affil{\small\em ETH Z\"urich}
\author {\textbf{\large Sergei Nayakshin}}
\affil{\small\em University of Leicester}
\author {\textbf{\large Jonathan J. Fortney}}
\affil{\small\em University of California, Santa Cruz}
\author {\textbf{\large Lucio Mayer}}
\affil{\small\em University of Z\"urich}
\author {\textbf{\large Yann Alibert }}
\affil{\small\em University of Bern}
\author {\textbf{\large Alan P. Boss}}
\affil{\small\em Carnegie Institution}

\begin{abstract}
\baselineskip = 11pt
\leftskip = 0.65in 
\rightskip = 0.65in
\parindent=1pc
{\small 
The large number of detected giant exoplanets offers the opportunity to improve our understanding of the formation mechanism, evolution, and interior structure of gas giant planets. The two main models for giant planet formation are {\it core accretion} and {\it disk instability}. 
There are substantial differences between these formation models, including formation timescale, favorable formation location, ideal disk properties for planetary formation, early evolution, planetary composition, etc. First, we summarize the two models including their substantial differences, advantages, and disadvantages, and suggest how theoretical models should be connected to available (and future) data. We next summarize current knowledge of the internal structures of solar- and extrasolar- giant planets. Finally, we suggest the next steps to be taken in giant planet exploration.  
 \\~\\~\\~}

\end{abstract}

\section{\textbf{INTRODUCTION}}

Giant planets play a critical role in shaping the architectures of planetary systems. Their large masses, orbital angular momentum, and fast formation make them prime candidates for driving rich dynamics among nascent planets, including exciting the orbits of small bodies and possibly delivering volatiles to terrestrial planets. 
Their bulk properties are also key for exploring the physical and chemical conditions of protoplanetary disks in which planets form. Furthermore, the diversity in properties (e.g. mass, radius, semi-major axis, density) of exoplanets calls for multiple formation and evolution mechanisms to be explored. 
As a result, detailed investigations of the formation mechanism, evolution, and interior structure of giant planets have been conducted for decades, although for most of that time, our understanding of giant planets was limited to the four outer planets in the Solar System. Naturally, theory was driven to explain the properties of those planets and their particular 
characteristics. The detections of giant planets outside the Solar System have led to the discovery of a rich and diverse population of giant exoplanets, many of which have measured radii and masses.  \par

Clues on the nature of giant planet formation might be revealed from the two 
giant planet correlations with stellar metallicity [Fe/H] of main sequence 
stars. The first is the correlation of the frequency of giant planets with 
stellar metallicity ({\it Gonzalez}, 1997; {\it Santos et al.}, 2004; 
{\it Fischer and Valenti}, 2005; {\it Mortier et al.}, 2013) which has now been determined for stellar masses 
from $\sim$0.3 to 2 M$_{\odot}$ and metallicities [Fe/H] between -1.0 and 0.5 
({\it Johnson et al.},~2010). Nevertheless, the robustness and generality of the correlation between giant planet occurrence and stellar metallicity is still under
investigation ({\it Maldonado et al.}, 2013). The second correlation suggests that the heavy element mass in giant planets is a function of stellar metallicity.  Metal-rich stars tend to have metal-rich planets. There is a large scatter, however, and planetary mass is a better predictor of metal enrichment than stellar metallicity (see section 5.2).
This explosion of new information demonstrates the need to understand planet formation in general, and presents an opportunity to test planet formation theories.  While the basic ideas of how giant planets are formed have not changed much, substantial progress has been made recently in highlighting the complexity of the models, in which effects that were perceived as second-order may actually be fundamental in shaping the distribution of giant planets. \par

Accurate measurements of the giant planets in the Solar System can provide valuable information on the internal structures of the planets, which are extremely important to test formation models.  We anticipate major progress in that direction when data from {\it Juno} and {\it Cassini Solstice} missions are available. In addition, the available information on the mean densities of giant exoplanets teaches us about the composition of giant planets in general.~This chapter summarizes the recent progress in giant planet studies.

\section{\textbf{GIANT PLANET FORMATION MODELS}}

In the following section we summarize the two formation models: 
{\it core accretion} ({\it CA}), the standard model, and {\it disk instability} ({\it DI}). 
In the {\it CA} model the formation of a giant planet begins with planetesimal coagulation and core formation, similar to terrestrial planets, which is then followed by accretion of a gaseous envelope. 
In the {\it DI} model, gas giant planets form as a result of gravitational fragmentation in the disk surrounding the young star.

\subsection{\textbf{CORE ACCRETION}}

The initial step in the formation of giant planets by {\it CA} 
is the same as that for terrestrial planets, namely the buildup of planetesimals with sizes from a few tens of meters to a few hundred 
kilometers. Planetesimal formation is itself undergoing current research and is addressed 
in the chapter of {\it Johansen et al.} Here we focus on the 
later stages, starting at the point where runaway growth of the
planetesimals has produced a planetary embryo having a fraction of an
Earth mass (M$_\oplus$), still surrounded by a swarm of planetesimals. The planetesimals
accrete onto the embryo to form the  heavy-element core of the giant
planet.  Once the core reaches a fraction of M$_\oplus$ and its escape velocity exceeds the local thermal speed of the gas in the disk, 
the core can begin to capture gas from the surrounding disk. Solids and gas then accrete concurrently. When the mass of gas ($M_\mathrm{env}$) approaches 
that of the core ($M_\mathrm{core}$), gas accretion can become very rapid and
can build the planet up to the Jupiter mass range. Gas accretion is cut
off at some point, either because of the dissipation of the disk and/or
by the gravitational interaction of the
planet with the disk, which tends to open up a gap in the planet's vicinity as described below.\par

\bigskip
\subsubsection{\textbf{Physics of the core-accretion model}}

\subsubsection*{Core accretion rate}
{\it CA} calculations generally start with an embryo of $\sim$ 0.01-0.1 M$_\oplus$ surrounded by  solid planetesimals, which would be expected to have a range of sizes.
Safronov's equation ({\it Safronov}, 1969) is a useful approximation for
the planetesimal accretion rate onto an embryo:
\begin{equation}
{{dM_\mathrm{solid}} \over {dt}}  =  \dot M_\mathrm{core} =
 \pi R_\mathrm{capt}^2
 \sigma_s \Omega  F_g,
\label{eq:accrete}
\end{equation} 
where $\pi R_\mathrm{capt}^2$ is the geometrical capture cross section, $\Omega$ is the
 orbital frequency, $\sigma_s$
is the disk's solid surface density, and $F_g$ is the gravitational
enhancement factor. If no gas is present, then $ R_\mathrm{capt} = R_\mathrm{core}$. In 
the presence of an envelope, gas drag and ablation can result in 
planetesimal deposition outside the core, with $ R_\mathrm{capt} > 
R_\mathrm{core}$.   

In the two-body approximation in which the
tidal effects of the central star are not included, $F_g$  is given by 
\begin{equation}
F_g = \left[1 + \left({{v_e} \over
{v}}\right)^2 \right],
\label{eq:fg}
\end{equation}
where $v_e$ is the escape velocity
from the embryo, and $v$ is the relative velocity of the embryo and
the accreting planetesimal.
 This approximation is valid only when the effect of the star is negligible, 
otherwise three-body effects 
must be taken into account, and the determination of $F_g$ is much more
complicated ({\it Greenzweig and Lissauer}, 1992). 

\subsubsection*{The disk of planetesimals}
The gravitational focussing factor depends strongly
on the excitation state of planetesimals (eccentricity and 
inclination), which itself is 
the result of excitation by the forming planet (and other planetesimals) and damping by gas drag.
Therefore, the value of $F_g$ must be determined (numerically)
by computing the planet's growth, the structure of the gas disk,
and the characteristics of planetesimals (dynamics and mass function) self-consistently. 
Gravitational interactions among planetesimals and,
   more importantly, the influence of the planetary
   embryo tend to stir up the velocities of the
   planetesimals, reducing $F_g$. {\it Fortier et al.}~(2013)
   show that giant planet formation is suppressed when this effect is included. 
   
   The large number of planetesimals in the feeding zone
   ($>10^{10}$) requires that the evolution, growth, and
   accretion rate of the planetesimals onto the embryo be
   handled in a statistical manner  ({\it Inaba et al.}, 2003;
   {\it Weidenschilling}, 2008, 2011; {\it Bromley and Kenyon}, 2011; {\it Kobayashi et al.}~2010).
   Planetesimals are distributed into a finite number of
   mass bins, with masses ranging from tens of meters to tens
   of kilometers. The planetesimals are also distributed
   into radial distance zones. The codes calculate the
   evolution of the size distribution and the orbital
   element distribution, as well as the accretion rate of
   each size onto the planetary embryo. The size distribution
   evolves through collisions, and, if relative velocities
   between planetesimals become large enough, through
   fragmentation.  The velocity distributions are affected by
   viscous stirring, dynamical friction, gravitational
   perturbations between non-overlapping orbits, and damping
   by gas drag from the disk. The outcome of fragmentation
   depends on specific impact energies, material strengths,
   and gravitational binding energies of the objects.
   Also to be included are migration of planetesimals through
   the disk, and interactions between the protoplanetary
   envelope and the various-sized planetesimals (see below). These
   simulations give $\dot M_\mathrm{core}$ as time progresses.

   Including all of these physical processes in planet
   formation models is numerically challenging, and all models
   involve some degree of approximation. The full
   problem of the self-consistent buildup to Jupiter mass,
   including the evolution of the planet, the gas disk, and
   the planetesimal distribution, has not been completed to date.
   
\subsubsection*{Envelope accretion rate}
The outer radius of the planet cannot be larger than  the 
smaller of the \emph{Bondi radius} 
$R_B$ or the \emph{Hill radius}  $R_H$.
Here $R_B = \frac{GM_p}{c_s^2}$ and
$R_H= a \left(\frac{M_p}{3 M_{\star}}\right)^{1/3}$,
where $M_p$ and $M_{\star}$ are the masses of the planet and star, 
respectively, $a$ is the distance of  the planet from  the star,
and $c_s$ is the sound speed in the disk.
However, three-dimensional
numerical simulations of the gas flow around a planet embedded in a
disk ({\it Lissauer et al.}, 2009) show that not all of the gas flowing through
the Hill volume is actually accreted by the planet. Much of it
just flows through the Hill volume and back out. A modified Hill
radius, within which gas does get accreted onto the planet, is found
to be $\sim 0.25 R_H$. Thus one can define an effective outer
radius for the planet:
\begin{equation}
R_{\rm eff} = \frac{GM_p}{c_s^2 + \frac{GM_p}{0.25 R_H}}~.
\label{eq:reff}
\end{equation}
During  the earlier phases of the evolution, when  $M_\mathrm{env}$ is
less than or comparable to $M_\mathrm{core}$, the gas accretion rate is
determined by the requirement that the outer planet radius 
$R_p = R_{\rm eff}$. As the envelope
contracts and $R_{\rm eff}$ expands, gas is added at the outer 
boundary to maintain this condition.

Once the planet has entered the phase of rapid gas accretion, 
the envelope is relatively massive, and it can contract rapidly. 
At some point the mass addition rate required by
the above condition exceeds the rate at which matter can be
supplied by the disk. The disk-limited rates are calculated 
from three-dimensional simulations of the flow around a planet
embedded in a disk ({\it Lubow and D'Angelo}, 2006; {\it Lissauer et al.}, 2009; {\it D'Angelo et al.}, 2011; {\it Bodenheimer et al.}, 2013; {\it Uribe
et al.}, 2013). The rates 
are eventually limited as the mass of the planet
grows, exerts tidal torques on the disk, and opens up a gap.
Although gas can still accrete through the gap, the rate becomes
slower as the gap width increases. 
Planet-disk interactions, however, can change the eccentricities and 
increase $\dot M_{\rm env}$  substantially even after gap formation 
({\it Kley and Dirksen}, 2006). 

\subsubsection*{Structure and evolution of the envelope}

The envelope of the forming planets is typically considered to consist mainly of hydrogen and helium (H, He), with a small fraction of heavy elements (high-Z material). 
The $Z$-component   of the planetary envelope can be stellar, sub-stellar, 
or super-stellar. 
If all the planetesimals reach the core without depositing mass in the envelope, the planetary envelope is likely 
to have a composition similar to the gas in the protoplanetary disk. 
However, enrichment of a gas giant through {\it CA} is not 
necessarily automatic, since the accreted gas could be depleted in solids 
as a result of planetesimal formation.  As a result, the accreted gas can be 
sub-stellar.  On the other hand, if planetesimals suffer a strong
mass ablation during their journey towards the core, the planetary envelope will be substantially enriched with heavy elements compared
to the disk gas, in particular during the first phases of planetary growth, when the solid
accretion rate can be much larger than the gas accretion rate.

Typically, {\it CA} models assume that the planetary envelope is spherically symmetric,
a good assumption, up to the phase of rapid gas accretion. The stellar structure equations of
hydrostatic equilibrium, energy transport by radiation or convection, and
energy conservation are used to determine its structure. During phases when
$R_p = R_\mathrm{eff}$, the density and temperature at the outer boundary
are set to the disk values $\rho_\mathrm{neb}$ and $T_\mathrm{neb}$, 
respectively.

The differential equations of stellar structure are then supplemented with the equation of
state (EOS) and the opacity. 
The grains in the envelope 
are typically assumed to have an interstellar size distribution. 
However {\it Podolak} (2003) shows that when grain coagulation and sedimentation in
the envelope are considered, the actual grain opacities, while near
interstellar values near the surface of the protoplanet, can drop by
orders of magnitude, relative to interstellar, deeper in the envelope.

\subsubsection*{Interaction of planetesimals with the envelope}
\label{sect:int}

The accreted planetesimals are
affected by gas drag,  which can result in a considerable enhancement of
the effective cross section for their capture, larger than 
$\pi R_\mathrm{core}^2$. The deposition of planetesimal material also has 
significant effects on the composition,  mean molecular
weight, opacity,  and EOS of the envelope. 
Individual orbits of planetesimals that pass through
the envelope are calculated ({\it Podolak et al.}, 1988). 
A variety of initial impact parameters
are chosen, and the critical case is found inside of which the
planetesimal is captured. Then $R_\mathrm{capt}$ 
is the pericenter of the critical
orbit. Even if $M_\mathrm{env} <0.01$ M$_\oplus$, this
effect can be significant, especially for smaller planetesimals
that are easily captured. The value of
$R_\mathrm{capt}$  can be up to 10 times $R_\mathrm{core}$, depending on 
$M_\mathrm{env}$  and  the planetesimal size.

The determination of the planetesimal mass ablation is in turn a complex process, and depends
strongly on the (poorly known) characteristics of planetesimals, among them their mass 
and mechanical properties. Low-mass planetesimals and/or low tensile 
strength planetesimals are likely to suffer strong mass ablation. 
These properties depend on the (yet unknown) planetesimal formation process (see chapter by {\it Johansen et al.}) and their interaction with forming planets and gas in the disk. 
Interestingly enough, the enrichment
of planetary envelopes by incoming planetesimals has a strong influence on the planetary growth by gas accretion itself.
In particular, as shown by {\it Hori and Ikoma} (2011), envelope pollution can 
strongly reduce the critical mass -- that is, the core mass required to
initiate rapid gas accretion.  This can speed up the formation of gas giant planets leading to objects with  
small cores and enriched envelopes at the end of the formation process.

On the other hand, it is often assumed that the ablated planetesimal
material eventually sinks to the core, releasing gravitational energy in
the process. Detailed calculations ({\it Iaroslavitz and Podolak}, 2007) show that
for particular envelope models the rocky or organic components do sink,
while the ices remain dissolved in the envelope. 

\bigskip
\subsubsection{\textbf{Phases of planetary formation}}

In classical {\it CA} models (i.e., without migration and disk
evolution) the important parameters for defining the formation of a giant planet  
are $\sigma_s$, $M_\star$ and $a$. 
These three quantities determine 
the \emph{isolation mass},  $M_\mathrm{iso} = 
\frac{8}{\sqrt{3}}(\pi C)^{3/2} M_\star^{-1/2} \sigma_s^{3/2} a^3$,  
which gives the maximum
 mass to which the embryo can grow, at which point it has accreted all
of the planetesimals in the feeding zone. $C$ is the number of Hill-sphere radii on each side of the
planet that is included in the feeding zone ({\it Pollack et al.}, 1996). 
 $M_\mathrm{iso}$ must roughly exceed
3 M$_\oplus$ in order to get mass accretion up to the Jupiter-mass
range during the lifetime of the disk. Although it is commonly stated that a core of $\approx$ 10 M$_\oplus$ is necessary to initiate rapid gas accretion the critical core mass can be both larger and smaller than this standard value, depending on the disk's properties, the physics included in the planetary formation process, and the core accretion rate. 
Additional parameters include
the gas-to-solid ratio in the disk, which for solar composition is
of order 100, but which decreases as the metallicity of the star
increases. The gas surface density determines $\rho_\mathrm{neb}$ and
a disk model determines $T_\mathrm{neb}$. The size distribution and
composition of the planetesimals in the disk are important for
the calculation of $\dot M_\mathrm{core}$.  Many calculations are
simplified in that a single-size planetesimal is used, with radius 
of sub-km in size and up to 100 km.
                    
The following are the main phases associated with the buildup of a giant planet in the {\it CA} model:
\begin{itemize}
\item 
Phase 1: Primary core accretion phase. The core accretes 
planetesimals until it has 
obtained most
of the solid mass within its gravitational reach, that is, its mass approaches
the isolation mass. $M_\mathrm{env}$ grows also, but it
remains only a tiny fraction of $M_\mathrm{core}$.

\item Phase 2: Slow envelope accretion.  The
$\dot M_\mathrm{core}$
drops considerably, and $\dot M_\mathrm{env}$ increases until it 
exceeds  $\dot M_\mathrm{core}$.
As $M_\mathrm{env}$ increases, the expansion of the zone of
gravitational influence allows more planetesimals to be accreted, but at a
slow, nearly constant rate. $\dot M_\mathrm{env}$ is limited by 
the ability
of the envelope to radiate energy and contract, so this rate is also slow,
about 2--3 times that of the core. Thus the time scale depends on
the opacity in the envelope, and if grain settling is incorporated into
the calculations, it falls in the range  0.5 to a few Myr.

\item Phase 3: Rapid gas accretion.
After  crossover ($M_\mathrm{core} = M_\mathrm{env}
 \approx \sqrt{2} M_\mathrm{iso}$), the rate of gas accretion  increases 
continuously, although the structure remains in quasi-hydrostatic equilibrium. $\dot M_\mathrm{env}$
greatly exceeds that of the core. Eventually the point is reached where the
protoplanetary disk cannot supply gas fast enough to keep up with the 
contraction of the envelope. $\dot M_\mathrm{env}$, instead of being
determined by the radiation of energy by the planet, is then determined
by the properties of the disk in the vicinity of the planet. Beyond this point, 
the accretion flow of the gas is hydrodynamic, nearly in free fall, onto a
protoplanet which is in hydrostatic equilibrium at a radius much smaller
than $R_\mathrm{eff}$.   The core mass
remains nearly constant on the time scale of the growth of the envelope.
The time scale to build up to $\approx$ 1 Jupiter mass (M$_\mathrm{Jup}$) is 
$10^4$-$10^6$ yr, depending on the assumed disk viscosity.

During disk-limited accretion, the gas envelope may be considered to consist of a hydrostatic
contracting  inner region, containing almost all of the mass, plus
a hydrodynamic accretion flow of gas onto it from the disk. The flow
will be supersonic, so the boundary condition at the outer edge of the
hydrostatic region will be determined by the properties of the shock
that forms at that interface ({\it Bodenheimer et al.}, 2000). Interestingly enough,
the properties of the shock have strong implications for the entropy of the gas that is finally 
accreted by the forming planet. Depending on the fraction of the incoming energy that is radiated
away through the shock, the entropy of accreted gas could differ substantially. This in
turn governs the evolution of the post-formation luminosity
({\it Mordasini et al.},
2012b), a quantity that could be accessible by direct imaging observations of planets. 

\item Phase 4: Contraction and cooling. 
The planet evolves at constant
mass. The main energy source is 
slow contraction in hydrostatic equilibrium, while an energy sink
arises from cooling by radiation. Stellar radiation heats up the surface.
Once accretion terminates, there is no longer an external energy source 
from planetesimal accretion,
but for a planet very close to its star, tidal dissipation and
magnetic dissipation may provide additional energy. If the
planet accretes a mass of $\approx $13 M$_\mathrm{Jup}$, deuterium
burning can set in, providing a significant energy supply. 
\end{itemize}

An example of a calculation of the first three phases is shown in Fig.~1. The first luminosity peak represents 
the maximum of $\dot M_\mathrm{core}$ during Phase 1, while the second peak corresponds to the maximum disk-limited accretion rate of gas during Phase 3. 
These three phases can change when the exchange of angular momentum between the forming planet and the disk is considered (see chapter by {\it Baruteau et al.}).  When the orbital parameters
of a forming planet change with time, its formation is clearly affected. 
First, the planet can form more rapidly (all other parameters being equal), and second,
its content in heavy elements is expected to be higher. It should be noted that the resulting higher content
in heavy elements, does not imply a higher core mass, and does not require a starting point in a
massive and/or very metal-rich disk.

\begin{figure}[ht!]
\begin{center}
\includegraphics[angle=0,height=68mm]{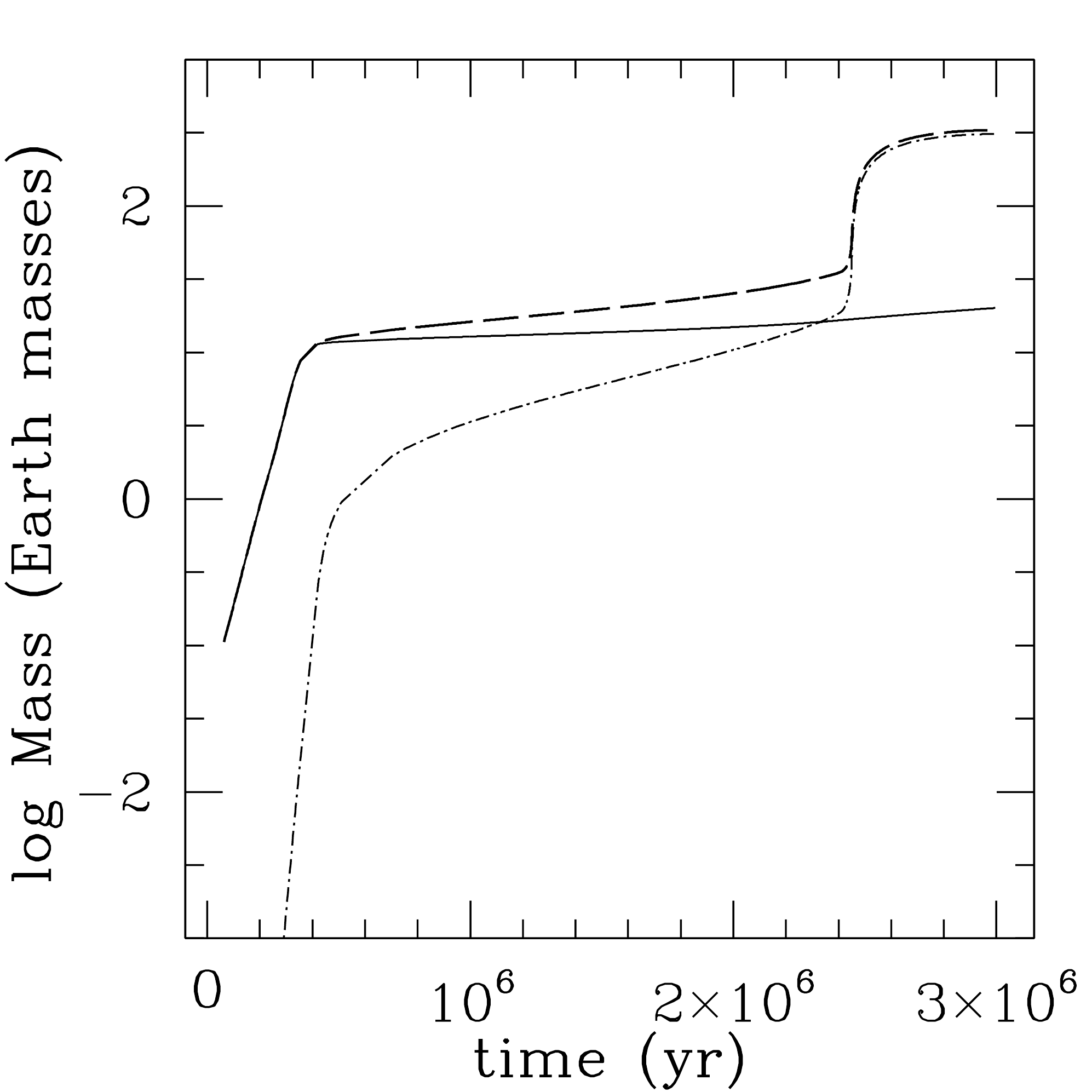}
\includegraphics[angle=0,height=68mm]{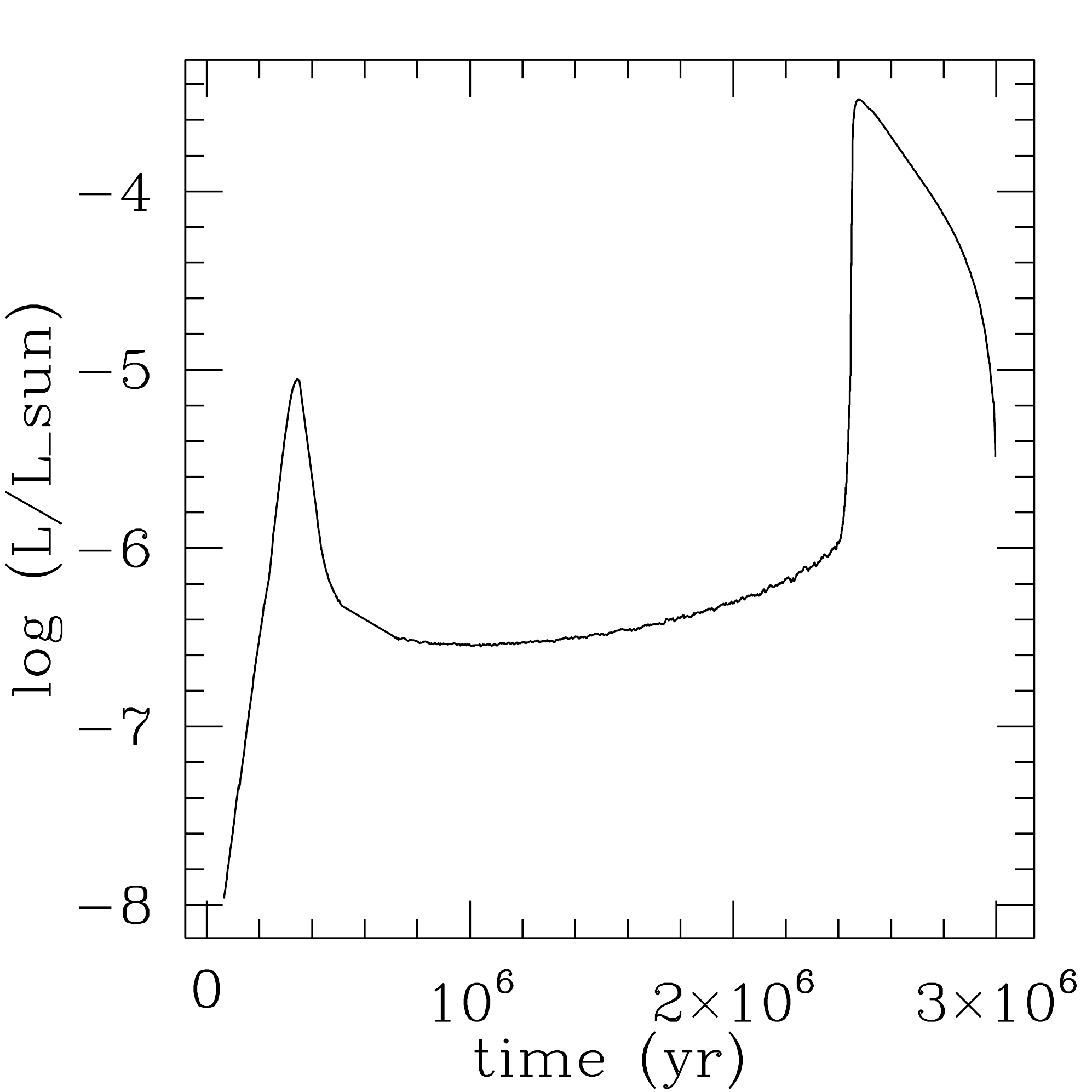}
\caption{
{\small
{\bf Top:} An example of the growth of a protoplanet (at 5.2 AU 
with $\sigma_s = 10 $ g cm$^{-2}$) 
by {\it CA}. The planet mass is shown vs.~ time. 
 \textit{Solid line}: core mass.
\textit{Dash-dot line}: envelope mass. \textit{Dashed line}: total mass. 
Grain opacities are reduced by a factor 50 from interstellar values.
Crossover mass is reached at 2.3 Myr, and 1 M$_\textrm{Jup}$ at
3 Myr.
{\bf Bottom:} The radiated luminosity vs.~time.  Adapted from {\it Lissauer et al.} (2009).
}
}
\label{fig:phasel}
\end{center}
\end{figure}

{\it Alibert et al.} (2004) have investigated how orbital migration of a forming planet,
whatever its origin, prevents the planet from reaching isolation. The same effect can also appear as a result
of the orbital drift of planetesimals by gas drag. As a consequence, phase 2 as presented above
is suppressed, and a forming planet would transit directly from phase 1 to phase 3. In that case the total formation timescale 
is substantially reduced. This process is illustrated in 
Fig.~2 which shows the formation of a Jupiter-mass planet for two
models which are identical except that one assumes {\it in situ} formation (dashed lines), while second includes 
planetary migration (solid lines). In the migrating case, the initial planetary location is 8 AU with $\sigma_s$ = $3.2$ g cm$^{-2}$, and the
planet migrates down to 5 AU, where the {\it in situ} model is computed with $\sigma_s$ = $7.5$ g cm$^{-2}$.
For the {\it in situ model}, the cross-over mass is reached after $\sim$ 50 Myr while in the migrating model,
cross-over mass is reached in less than a million years. One should also note that when migration is included the total amount of heavy elements 
is increased by a factor $\sim 2$. Both models are computed assuming the same disk mass, which is around three
times the Minimum Mass Solar Nebula (MMSN), compatible with disk-mass estimations 
from observations (e.g. {\it Andrews et al.} 2011; see Section 2.1.4).

\begin{figure}[h!]
\begin{center}
\includegraphics[angle=0,height=70mm]{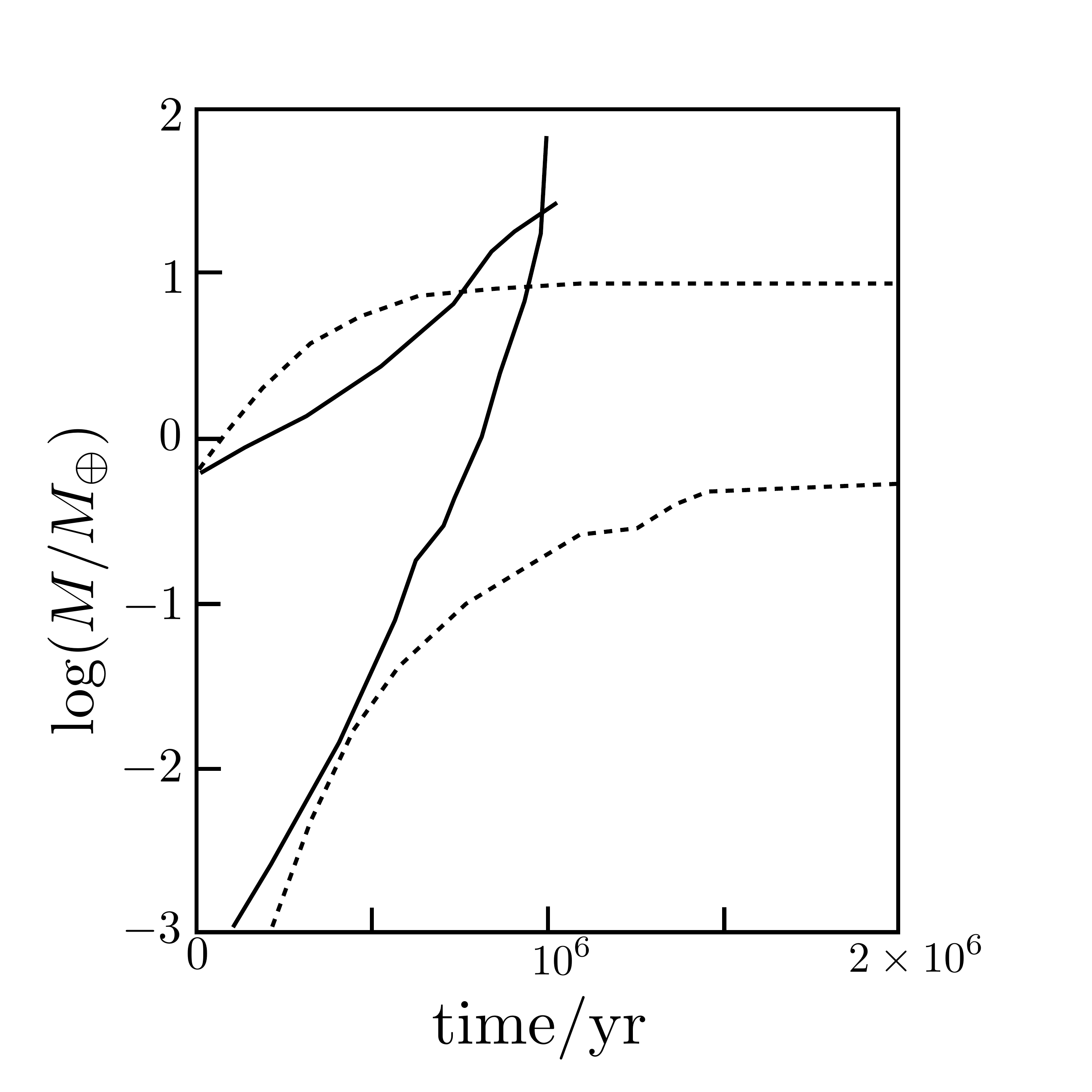}
\end{center}
\caption{
{\small
The total mass of heavy elements (core+envelope) and gaseous envelope (H/He) mass vs.~time until cross-over mass is reached. 
{\it Solid lines}: a migrating planet starting at 8 AU, 
without gap formation; {\it Dashed lines}: {\it in situ} formation. 
Adapted from {\it Alibert et al.} (2005a).
}
}
\label{masses}
\end{figure}

\bigskip
\subsubsection{\textbf{Termination of accretion}}
\label{sect:gasd}

Two mechanisms have been discussed for the termination of the rapid gas accretion
onto a protoplanet: disk dissipation and gap opening. The general picture regarding
the evolution of a disk involves a gradual reduction in disk mass and disk
gas density over a period of 2--4 Myr, after which a relatively rapid
(a few $10^5$ yr) clearing phase occurs. If the giant planet fully forms before
the clearing phase occurs, then the disk dissipation mechanism is not
relevant. 
 However this mechanism could work to explain the structure of Uranus and Neptune 
({\it Pollack et al.}, 1996). The core accretion time for these planets at their
present orbital positions is so long that it is very unlikely that they
formed there. It has been suggested (e.g., {\it Tsiganis et al.}, 2005) that these
   planets formed at closer distances, between 5 and 20 AU, and
   were later scattered outward. 
Even at these distances
the formation time for the core could be long enough so that
substantial disk dissipation occurs during Phase 2, and the planet never reaches the crossover mass. 

For planets that reach the phase of rapid gas accretion it seems as
though gap opening is the critical factor.
Once the forming planet reaches a sufficient mass, its gravitational
influence on the disk results in a torque on disk material interior
to the planet's orbit; the torque removes angular momentum from the
material and allows it to move closer to the star.
At the same time a torque is also exerted on the material external to
the planet; this material gains angular momentum and is forced to move
to larger distances. In this way a gap, or at least a disk region of
reduced density, tends to form near the planet's orbit ({\it Lin and Papaloizou}, 1979).
However, there is an opposing effect: the viscous frictional torque
in the differentially rotating disk tends to transfer angular
momentum outwards and mass inwards; this effect tends
to fill in the gap. 

The condition for a gap to open then requires
that the gravitational effect of the planet dominates over the 
viscous and pressure effects ({\it Crida et al.}, 2006). If $R_H > H$ (the disk scale height), which is the
normal situation for the appropriate mass range, then the 
minimum planet mass  $M_p$ for gap-opening is ({\it D'Angelo et al.}, 2011)
\begin{equation}
\left(\frac{M_p}{M_\star}\right)^2 \approx 3 \pi \alpha f 
  \left(\frac{H}{a}\right)^2\left(\frac{R_H}{a}\right)^3,
\end{equation}
where $f$ is a parameter of order unity, 
and $\alpha$ is the disk viscosity parameter.
For $H/a = 0.05$ and $\alpha=10^{-2}$ the minimum estimated mass
is about 0.2 M$_\mathrm{Jup}$ around a solar-mass star. 
Numerical simulations of disk evolution 
with an embedded planet of various masses show that by the time
this mass is reached, the gas density near the orbit has been
reduced to about 60\% of the unperturbed value; at  1.0 M$_\mathrm{Jup}$
the reduction is over 95\% ({\it D'Angelo et al.}, 2011).

\begin{figure}[h!]
\begin{center}
\includegraphics[angle=0,height=65mm]{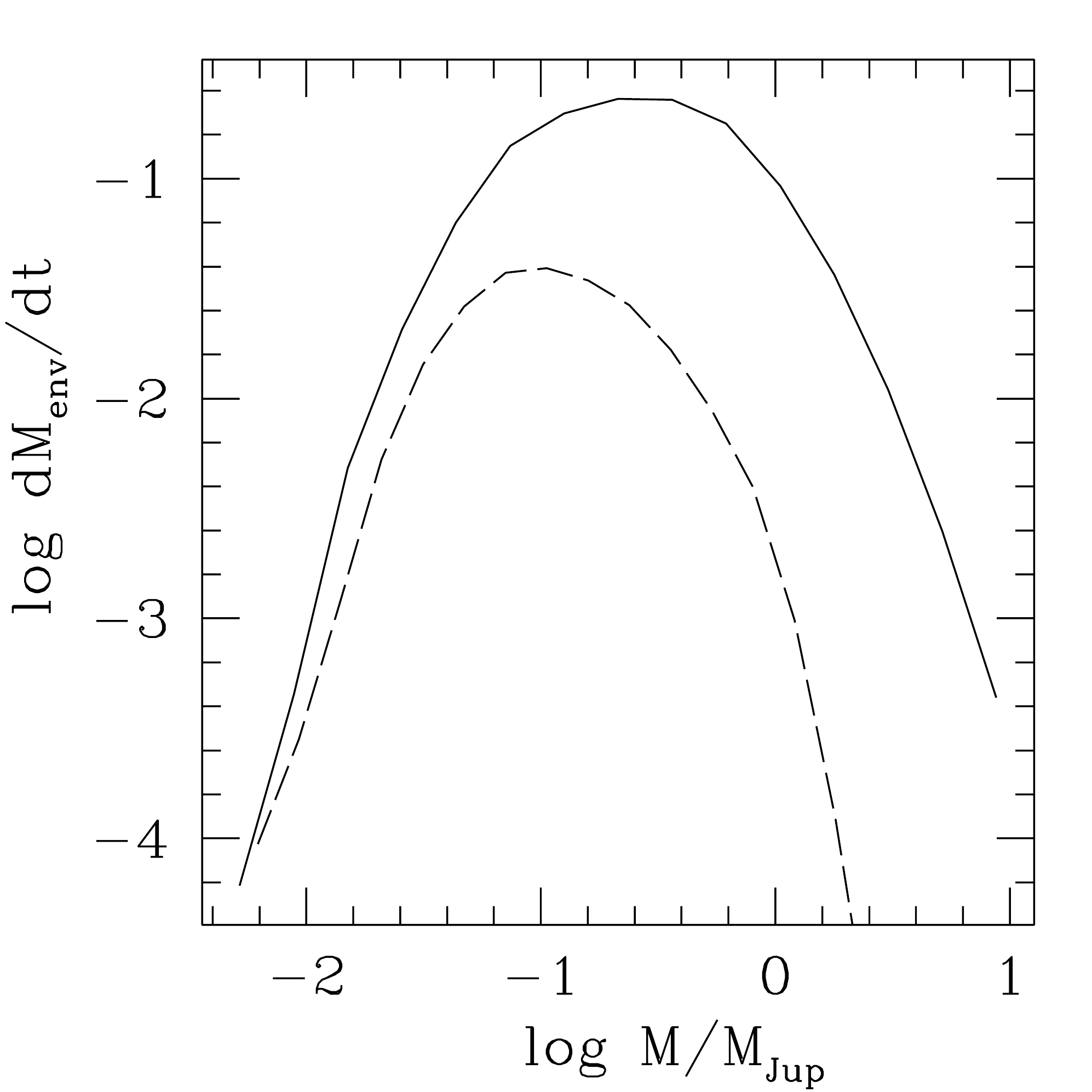}
\caption{
{\small 
Disk-limited gas accretion rates (M$_{\oplus}$/yr) vs.~total planetary mass for a planet 
at 5.2 AU around a 1 M$_{\odot}$ star in a disk with initial gas density $\sigma_g$=700 g cm$^{-2}$.
(\textit{Solid line}): disk viscosity $\alpha=4\times10^{-3}$.
(\textit{Dashed line}):  disk viscosity $\alpha=4\times10^{-4}$. 
At masses below $\sim$0.25 M$_\mathrm{Jup}$ the actual rates fall below
these curves because they are limited by the rate at which the gaseous
envelope contracts. These rates scale as $\sigma_g a^2 \Omega$, where 
$a$ is the orbital distance, and $\Omega$ is the orbital
frequency. Adapted from {\it Lissauer et al.} (2009).
}
}
\label{fig:genn}
\end{center}
\end{figure}

Even after gap opening has been initiated, gas can still flow through it and
accrete onto the planet, although the rate of accretion decreases as
the gap becomes deeper and wider. Fig.~ 
\ref{fig:genn} shows that, depending on the viscosity, the peak in 
$\dot M_\mathrm{env}$ occurs for a planet of mass 0.1 to 1 M$_\textrm{Jup}$
around a solar-mass star. However beyond that point, gas continues to
be accreted 
through the gap until its density is reduced by a factor 1000,
leading to a final planet mass of up to 5--10 M$_\mathrm{Jup}$ for $\alpha \approx
4 \times10^{-3}$.
Further accretion is possible if the protoplanet has an
eccentric orbit.  Thus the
maximum mass of a planet  
appears to be close to the upper limit of the 
 observed mass range for exoplanets.  Planets which end up
in the Jupiter-Saturn mass range, then, 
would  have to be explained by formation in a disk region with 
cold temperatures (low $H/a$) and low
viscosity, or at least in a disk which has evolved to such conditions at
the time of rapid gas accretion. 
Angular momentum is expected to be transferred from the disk to the 
planet during the gap-opening process,
leading to a sub-disk around the planetary core ({\it Ayliffe and Bate}, 2012). One might
think that the presence of this sub-disk would slow down the accretion process onto
the planet; this effect must still be characterized, taking into account accretion through the planet's polar regions. 
In any  case, the limiting gas accretion rate depends, in particular,
on the disk viscosity and gas surface density. The latter quantity
is known to evolve on timescales comparable to that  of buildup of planetary
cores (a few Myr). As a consequence, at a time when the critical mass, i.e. the core mass needed to initiate rapid gas accretion (e.g. {\it Mizuno}, 1980), is reached, the limiting accretion rate
has decreased by up to two orders of magnitude, compared to a 
classical young disk, reaching values 
$\sim 10^{-4} \mearth$/yr or lower. Under these circumstances, 
the accretion of 1 M$_\mathrm{Jup}$ takes $\approx$  
3 Myr, comparable with disk lifetimes. 
Such calculations have been used in order to explain the heavy element mass in Jupiter and Saturn  
({\it Alibert et al.} 2005b), as well as the observed mass function of exoplanets ({\it Mordasini et al.} 2009a,b).
An example is the model shown in Fig. 1, which reached a final mass of 
 1 M$_\mathrm{Jup}$ without any assumed cutoff, in a disk with viscosity
parameter $\alpha=4 \times 10^{-4}$ and a disk lifetime of 3 Myr ({\it 
Lissauer et al.}, 2009).

\bigskip
\subsubsection{\textbf{Core accretion as a function of parameters}}

\emph{Effect of position in disk}.
The presence of exoplanets near their stars has led to the suggestion
({\it Lin et al.}, 1996) that giant planets form in the 5--10 AU region according to the
picture just discussed, but during the formation process they migrate inwards.
However one must also consider the possibility that these planets could 
actually form at  or near their present locations.  In the very inner
regions of disks, $\sim$0.05 AU where the hot Jupiters are found, there are
several difficulties. First, the temperatures are high so there is very
little condensible material.~Second, the typical mass distribution in the inner disk does not provide sufficient material to form 5--10 M$_\oplus$ cores, necessary
for giant planet formation. Third, even if a planet did form, it would very
likely migrate into the star (though outward migration may be
  feasible for low mass planets; {\it Kley et al.~2009; Hellary and 
    Nelson 2012}). A possible way of circumventing these
problems is to build up the core by migration  of planetesimals or small
solid objects inward through the disk  and then collecting them in the inner
regions of the disk 
(e.g., {\it Ward}, 1997; {\it Bodenheimer et al.}, 2000).  
Also, planet formation is more likely in a metal-rich or massive inner disk.
The third difficulty can be
averted if one assumes that the inner disk has been cleared, 
for example by the stellar magnetic field, so that the torques on the 
planet would be negligible. 

The possibility of migration introduces an additional complexity. In this case, it is difficult to define the formation location of a
planet, since they can migrate over very large distances,
experiencing very diverse conditions in different parts of the disk.
Nonetheless, the following facts are still valid: (1)  the solid accretion rate decreases as a 
function of distance,  and (2) the amount of available solid material decreases
as one moves inward toward the star. If there is no significant migration, the optimum
location for core-accretion formation of a giant planet is between 5 and 10 AU for a solar-mass star. 
The maximum distance at which a giant planet can form in a minimum-mass solar nebula is 40--50 AU
  ({\it Rafikov}, 2011), based on the assumption of a maximum rate of planetesimal accretion onto the core. \par

\emph{The effect of the mass of the star}.       
Typical observed disk masses decrease roughly linearly with the mass of the central star from $M_{\star}$ of a few solar masses down to less than 0.1 $M_{\odot}$ 
    with a typical $M_{disk}/M_{\star}$= 0.002-0.006 ({\it Andrews et al.}, 2013).
However,  observed disk masses show a large spread at a given
stellar mass, from  0.001 M$_\odot$ to 0.1 M$_\odot$ for a 1 M$_\odot$ star. 
Furthermore, disk masses are uncertain because they are determined from
dust continuum observations, which do not measure the gas mass and which
are affected by uncertainty in the dust opacity. 
Characteristic disk lifetimes 
extend from a median of 3 Myr to a maximum of 10 Myr ({\it Hillenbrand}, 2008). 
Separating out the disk
dispersal times as a function of  mass  has proved to be difficult
({\it Kennedy and Kenyon}, 2009), with no strong dependence observed. As a useful first
approximation, one can say that disk lifetimes are independent of stellar
mass for $ M _\star < $1.5  M$_\odot$, but there is a decrease in lifetimes
for $ M_\star  > $1.5 M$_\odot$, possibly because of more effective dissipation
by ultraviolet and X-ray irradiation (see chapter by {\it Alexander
  et al.}). Again, for a given stellar mass, there
is a range of disk lifetimes, from about 1 to 10 Myr.

If the disk mass in solids scales with stellar mass, then the solid
surface density in disks around stars less massive than the Sun is
expected to be low, leading to slower formation. 
In addition, 
the dynamical time $\Omega^{-1}$ is longer,
resulting in  longer accretion times. 
Thus it should be more difficult to form a Jupiter-like planet at           
a given  radial distance around a low-mass star than around a
solar-mass star ({\it Laughlin et al.}, 2004),
on a time scale comparable to the 
characteristic 2.5 Myr lifetime  of disks around low-mass stars ({\it Mamajek}, 2009). 
While the existence of such a disk is possible, the probability of forming the planet around the low-mass star is clearly smaller than for a  1 M$_\odot$ star.  
A similar argument, in the opposite sense,
 applies to stars that are more massive than the Sun. These qualitative considerations
are consistent with the observed correlation that the probability of
a star hosting a planet of $\sim 0.5$ M$_{\text{Jup}}$ or greater is
proportional to $M_\star$  ({\it Johnson et al.}, 2010). In addition, more detailed theoretical calculations going beyond these order of magnitude estimates, 
predict that giant planets are less frequent around low-mass stars ({\it Ida and Lin}, 2005; {\it Alibert et al.}, 2011). 
\par

\emph{Effect of the metallicity of the star}:
Clearly if the metallicity of the
star [Fe/H], increases, 
the metallicity of its disk also increases. If
the abundances in the grain-forming material, namely ices, organics
and silicates, increase in proportion to the iron abundance,
one gets a higher $\sigma_s$ at a given distance around a star of 
higher metallicity.  Indeed observations show that the
planet occurrence probability at a given iron abundance increases
with the stellar silicon abundance ({\it Robinson et al.}, 2006), consistent with this picture.
 The higher  $\sigma_s$ gives a shorter formation
time for the core at a given distance, with time inversely 
proportional to $\sqrt{\sigma_s}$. 
However there is an opposing
effect: the higher dust opacity in the protoplanet's envelope would
tend to lengthen the time for gas accretion. On the other hand the 
higher dust density speeds up the process of dust
coagulation and settling, which reduces the opacity, so the envelope
effect may not be very significant. In fact the observed probability of giant
planet formation increases faster than linearly with the
metallicity. A possible explanation is 
that a higher $M_\mathrm{iso}$ gives a higher luminosity during Phase 2 and a 
significantly reduced accretion time for the gas ({\it Ida and Lin}, 2004).
Thus there is an increased probability that
the giant planet can form before the disk dissipates. \par
Clearly there are other parameters which affect the {\it CA} process, 
including the disk mass (at a given stellar mass), the disk lifetime,
the disk viscosity, and the dust opacity in the protoplanetary
envelope. The first two of these effects, along with disk
metallicity,  have been studied by {\it Mordasini et al.} 2012a, while a detailed model of the last effect
was presented by {\it Movshovitz et al.} (2010).
  
\bigskip
\noindent
\subsection{\textbf{DISK INSTABILITY}}
\bigskip

Since PPV {\it DI} studies have made large advancements in
isolating conditions that are likely to lead to disk fragmentation i.e., the breakup of a protoplanetary disk into one or more self-gravitating clumps. 
Recent studies show that while the period of clump formation may be short-lived, fragments,
whether permanent or transient, can have a fundamental influence on the
subsequent evolution of a nascent stellar and/or planetary system. 
In this section, we summarize the {\it DI} model for giant planet formation.

\noindent
\subsubsection{\textbf{Disk Fragmentation}\label{sec:diskfrag_intro}}

The formation of giant planets in the {\it DI} model is conditioned by disk fragmentation. 
Disk perturbations will grow and form density enhancements throughout
regions of a disk wherever the destabilizing effects of self-gravity
become as important as, or dominate over the stabilizing effects of
pressure and shear. 
The threshold for the growth of axisymmetric density perturbations in a
thin gaseous disk is given by the Toomre (1964) criterion 
$Q= c_s \kappa/\pi G \sigma_g \sim Ê1$, where $c_s$ is the sound speed,
$\kappa$ is the epicyclic frequency, and $\sigma_g$ is the gas surface
density. The $Q$ criterion is strictly derived for an axisymmetric,
infinitesimally thin disk, and describes the disk's response at a
given radius to small (linear) axisymmetric perturbations. 
However, it is routinely applied to global, vertically stratified disks subject to non-axisymmetric perturbations with considerable degree of success.

Numerous numerical studies in 2D and 3D have demonstrated
that a disk will develop spiral structure for $Q\lesssim 1.7$, well
before $Q=1$ is reached ({\it Durisen et al.}~2007). 
These non-axisymmetric perturbations are the physical outcome of gravitational instabilities 
in realistic disks. They produce disk torques and shocks that
redistribute mass and angular momentum, and provide a source of heating throughout
gravitationally unstable regions.
Thus, spiral arms can act to stabilize the disk by increasing the
local sound speed and spreading the disk mass out.
In contrast, radiative cooling (with or without convection present)
will decrease the sound speed and destabilize the disk.

When the heating and mass transport from spiral arms can balance cooling and/or
infall of gas from the molecular cloud core when the disk is still young, 
persistent spiral structure can exist in the
disk.  In other cases, cooling and/or mass infall can lead to a second
instability, i.e., the collapse of regions of spiral arms into bound,
self-gravitating clumps. 
Determining whether or not clumps formed by gravitational instability can evolve into gas giant planets requires a combination of simulations that span several orders of magnitude in densities and spatial scales. 

The exact conditions that lead to a disk breaking up into self-gravitating clumps are still under investigation.   
The principal and firmly established condition is that the disk must be strongly self-gravitating, where $Q<1.4$  is necessary for an isothermal disk to fragment (e.g., {\it Nelson et al.}~1998, {\it Mayer et al.}~2004). 
This value is higher than the often assumed $Q<1$.  For non-isothermal disks, the $Q$ condition alone is insufficient, and a cooling time scale is usually used to identify regions that are expected to form clumps.   {\it Gammie} (2001) found that for 2D disks with an adiabatic index $\Gamma=2$, disks fragmented when $\beta<3$ where $\beta =  t_{cool} \Omega$ and $t_{cool}$ is the cooling timescale.  Here, $\Omega\approx \kappa$ is the local orbital frequency.  {\it Rice et  al.}~(2005) studied this cooling criterion in more detail and found that fragmentation occurred in 3D disks when $\beta \lesssim 6$ for $\gamma =5/3$ and $\beta \lesssim 12$  for $\gamma=7/5$.  
While there have been suggestions that the critical cooling timescales might be affected by the disk's thermal history ({\it Clarke et al.}, 2007), the temperature dependence of the cooling law ({\it Cossins et al.}, 2010) and the star and disk properties ({\it Meru and Bate}, 2011b), the $\beta$ condition for fragmentation was recently questioned anew by {\it Meru and Bate} (2011a), who found that the cooling time constraint was not converged with numerical resolution in SPH studies. Additional investigations of the topic were presented by {\it Lodato and Clarke} (2011) and {\it Paardekooper et al.} (2011). 

{\small
\begin{deluxetable}{cccccc}
\tablecaption{
{\small
Comparison of numerical techniques for smoothed 
particle hydrodynamics (SPH) codes: $\alpha_{SPH}$ and $\beta_{SPH}$
are coefficients of artificial viscosity, $N_{max}$ is the maximum
total number of SPH particles used to represent the disk, and
the disk thermodynamics is classified as either a simple $\beta$
cooling prescription, or more detailed, flux-limited radiative
transfer (RT). 
Key to references: (1) {\em Alexander et al.},
  2008 (2) {\em Meyer 
et al.}, 2007 (3) {\em Cha and Nayakshin}, 2011 (4) {\em Meru and 
Bate}, 2010 (5) {\em Meru and 
Bate}, 2012 (6) {\em Stamatellos and Whitworth}
}
\label{tbl-1}}
\tablehead{\colhead{\quad \quad code \quad \quad} & 
\colhead{\quad \quad $\alpha_{SPH}$ \quad \quad} &
\colhead{\quad \quad $\beta_{SPH}$ \quad \quad} &
\colhead{\quad \quad $N_{max}$ \quad \quad} &
\colhead{\quad \quad thermodynamics \quad \quad} &
\colhead{\quad \quad Ref. \quad \quad}
}
\startdata
GADGET2    & 1.0  & 2.0  & $1.3 \times 10^7$ & $\beta$ cooling & 1  \\
GASOLINE   & 1.0  & 2.0  & $1.0 \times 10^6$  & flux-limited RT & 2  \\
GADGET3    & 0.0  & 0.0  & $1.5 \times 10^6$    & $\beta$ cooling  & 3 \\
SPH/MPI    & 0.1  & 0.2  & $2.5 \times 10^5$  & flux-limited RT  & 4 \\
SPH/MPI    & 0.01-10  & 0.2-2.0  & $1.6 \times 10^7$  & $\beta$ cooling  & 5 \\
DRAGON     & 0.1  & 0.2  & $2.5 \times 10^5$  & flux-limited RT & 6  \\
\enddata
\end{deluxetable}
}

{\it Meru and Bate}
(2012) showed that the previous
non-convergent results were due to the effects of artificial viscosity
in SPH codes and suggested that the critical cooling timescale
may be at least as large as $\beta$ = 20 and possibly even as high as
$\beta$ = 30. However, convergence between different codes has not yet
been seen. Regardless, while the exact value might change the radial
location beyond which fragmentation can occur, there is a growing
consensus that fragmentation is more common at large than at small
disk radii. Moreover, if the disk is still accreting gas from the
molecular core, a high mass accretion rate onto the disk may become
larger than mass transport through the disk by spiral waves, leading
to disk fragmentation in disks that are cooling relatively slowly.  

The mass of the star can also affect disk fragmentation.  On one hand an increase in stellar mass will increase the Toomre parameter via the angular frequency.  On the other hand, more massive stars are thought to harbor more massive disks thus decreasing the Toomre parameter.  In addition, the temperature of different mass stars will also affect the disk temperatures and hence the corresponding Toomre parameter.  The combination of these effects suggests that disk fragmentation around different mass stars is not straightforward and has to be investigated carefully.  
{\it Boss} (2011) explored this effect and found that increasing the stellar mass resulted in an increase in the number of fragments. {\it Boss} (2006b) also showed that gravitational instability can offer a mechanism for giant planet formation around M dwarfs where {\it CA} often fails due to the long formation timescales in disks around low-mass stars. However, there is not yet a consensus on the conditions that lead to disk fragmentation.  For example, {\it Cai et al.} (2008) found that clumps are
not likely to form around low-mass stars. It is clear that follow-up studies are needed to determine how the efficiency of disk fragmentation varies with stellar mass. Moreover, clump evolution itself may be dependent on the host star's mass, further leading to observational differences.

 \noindent
\subsubsection{\textbf{Differences Between Numerical Models of {\it DI}}}

One of the major difficulties in understanding the conditions that
lead to clump formation is that fragmentation is inherently a highly
non-linear process.  While analytical models have been presented
(e.g., {\em Rafikov}, 2007; {\it Clarke}, 2009), theoretical efforts to model {\it DI} 
must ultimately be done numerically. 
In this section we restrict attention to numerical models, and in particular 
to models focused on whether {\it DI} can lead to the formation of self-gravitating
clumps, as opposed to other processes in {\it DI} models. Further details on 
the numerical methods for {\it DI} models can be found in {\em Durisen et al.} (2007). 
Here we concentrate on {\it DI} numerical models published since 
{\em PPV}.

Tables 1 and 2 list some of the primary differences between global numerical {\it DI}
models, with the former table listing Lagrangian, 3D smoothed particle hydrodynamics 
(SPH) codes, and the latter grid-based, finite-differences (FD) codes. The 
SPH codes include GADGET2 ({\em Alexander et al.}, 2008), GASOLINE ({\em Mayer 
et al.}, 2007), GADGET3 ({\em Cha and Nayakshin}, 2011), SPH/MPI ({\em Meru and 
Bate}, 2010, 2012), and DRAGON ({\em Stamatellos and Whitworth}, 2008).
The FD codes include FARGO ({\em Meru and Bate}, 2012; {\it Paardekooper et al.}, 2011), EDTONS ({\em Boss},2012), CHYMERA/BDNL ({\em Boley et al.}, 2007), IUHG ({\em Michael
and Durisen}, 2010), and ORION ({\em Kratter et al.}, 2010a), which is
an adaptive mesh refinement (AMR) code. 

Artificial viscosity (e.g., {\em Boss}, 2006a; {\em Pickett and Durisen}, 2007) 
and spatial resolution (e.g., {\em Nelson}, 2006; {\em Meru and Bate}, 2011a) 
have been shown repeatedly to be important factors in the accuracy of {\it DI}
numerical calculations and are also not mutually exclusive ({\em Meru and Bate}, 2012). 
For both FD and SPH codes, it is important to have 
sufficient spatial resolution to resolve the Jeans and Toomre length scales 
({\em Nelson}, 2006), otherwise numerically-derived fragmentation might result.
FD codes avoid this outcome by monitoring these criteria throughout the
grid, while SPH codes enforce a minimum number of nearest neighbor 
particles (typically 32 or 64), in order to minimize the loss of spatial 
resolution during an evolution. For SPH codes, some of the other key parameters 
are the smoothing length, $h$, which is variable in modern implementations, and 
the softening parameter used to derive the disk's self-gravity from that of the ensemble
of particles, the latter also having an analogous importance in grid codes ({\it M{\"u}ller et al.} 2012). Differences in the 
handling of radiative transfer can also lead to spurious results, prompting the testing
of numerical schemes against analytical results ({\em Boley et al.},
2007; {\em Boss}, 2009). Even indirect factors such as whether the central
protostar is allowed to wobble to preserve the center of mass of the star-disk
system may have an effect (cf., {\em Michael and Durisen}, 2010; {\em Boss}, 2012).

{\small
\begin{deluxetable}{cccccc}
\tablecaption{
{\small
Comparison of numerical techniques for finite differences
(FD) codes: dimensionality is the number of dimensions in the solution
(2D or 3D), $C_q$ is a coefficient of artificial viscosity, $N_{max}$ is the 
maximum total effective number of grid points (including midplane symmetry),
and the disk thermodynamics is classified as isothermal, simple $\beta$
cooling, or more detailed, flux-limited radiative transfer (RT).
Key to references: (1) {\em Meru and Bate}, 2012 (2) {\em Michael
and Durisen}, 2010 (3) {\em Boley et al.}, 2007 (4) {\em Boss}, 
2008, 2012 (5) {\em Kratter et al.}, 2010a
}
\label{tbl-2}}
\tablehead{\colhead{\quad \quad code \quad \quad} & 
\colhead{\quad \quad dimensionality \quad \quad} &
\colhead{\quad \quad $C_q$ \quad \quad} &
\colhead{\quad \quad $N_{max}$ \quad \quad} &
\colhead{\quad \quad thermodynamics \quad \quad} &
\colhead{\quad \quad Ref. \quad \quad}
}
\startdata
FARGO         &  2D   & 0-2.5      & $5 \times 10^7$    & $\beta$ cooling  & 1 \\
IUHG          &  3D   & 3.0      & $2 \times 10^6$    & $\beta$ cooling & 2  \\
CHYMERA/BDNL  &  3D   & 3.0      & $8 \times 10^6$    & flux-limited RT & 3  \\
EDTONS        &  3D   & 0.0      & $8 \times 10^6$    & flux-limited RT  & 4 \\
ORION(AMR)    &  3D   & $\ne 0$  & $3 \times 10^{14}$ & isothermal      & 5  \\
\enddata
\end{deluxetable}  
}

While SPH codes with spatially adaptive smoothing lengths are able to follow
arbitrarily high density clumps, most of the FD codes used to date have either
fixed or locally refined grids, which limits their abilities to follow
clumps in evolving disks. Pioneering {\it DI} calculations with the FLASH adaptive mesh
refinement (AMR) code have been shown to achieve very good convergence with
SPH simulations conducted with the SPH code GASOLINE, although only
simple thermodynamics with a prescribed equation of state was employed ({\it Mayer and Gawryszczak}, 2008).
A challenge for future work is to develop AMR simulations that retain the crucial physics of
the {\it DI} mechanism. 
\par

 \noindent
\subsubsection{Efficiency of Disk Instability With Stellar/Disk Metallicity}

While more giant planets have been discovered around metal-rich stars ({\it Fischer 
and Valenti} 2005, {\it Mayor et al.}, 2011) some planets have been discovered 
around metal-poor hosts (e.g., {\it Santos et al.}, 2011). 
Furthermore, disks around low 
metallicity stars are dispersed more rapidly (in $\lesssim$ 1 Myr; {\it Yasui 
et al.} 2009), compared to classical T Tauri disks which may survive for up to
10 Myr.  If planets form in such systems the process must be rapid.

Fragmentation is heavily dependent on disk thermodynamics,  
which is affected by the disk opacity and mean molecular weight.  Both 
of these scale directly with the disk metallicity.  With recent developments 
of complex radiative transfer models, one can try to understand the effects 
of metallicity via these parameters.  A decreased opacity is equivalent 
to simulating a lower metallicity environment or substantial grain growth. {\it Cai et al.} (2006) 
showed that gravitational instability becomes stronger when the opacity is 
reduced.  This is because the cooling is more rapid in an optically thin 
environment as the energy is able to stream out of the disk at a faster rate. 
Earlier studies investigating the effect of opacity on disk fragmentation 
suggested that varying it by a factor of 10 or even 50 had no effect on 
the fragmentation outcome ({\it Boss} 2002; {\it Mayer et al.}, 2007).  
These were simulations carried out in the inner disk ($\lesssim$ 30AU) where it is optically thick and so changes of this order did not significantly 
affect the cooling.  On the other hand, decreasing the opacity by two 
orders of magnitude allowed the inner disks to fragment when otherwise 
identical disks with higher Rosseland mean opacities did not ({\it Meru 
and Bate}, 2010).

At large radii big opacity changes are not  required for fragmentation 
to occur: decreasing the opacity by a factor of a few to one order of 
magnitude is sufficient to increase the cooling rate to allow such disks to 
fragment when otherwise they would not have done so with interstellar 
opacity values ({\it Meru and Bate}, 2010; {\it Rogers and Wadsley}, 2012).  
Nevertheless in general, when considering the effects of radiative transfer 
and stellar irradiation, opacity values that are lower than interstellar 
Rosseland mean values appear to be required.  These results suggest that 
in a lower metallicity disk, the critical radius beyond which the disk 
can fragment will move to smaller radii since more of the disk will be 
able to cool at the fast rate needed for fragmentation.
Alternatively, sudden opacity changes, occurring on extremely short timescales, could be induced by the
response of realistic composite dust grains to spiral shocks. Using post-processing
analysis of 3D disk simulations {\it Podolak et al.} (2011) have found that the ice shells of grains with refractory cores are rapidly lost
as they pass through the shock. This alters the grain size distribution as ice migrates between grains with different sizes, which results in a change in the opacity to several times lower than standard values. Since the opacity change would occur on timescales
much shorter than the orbital time,  cooling 
would be locally enhanced, possibly promoting fragmentation even in
the inner disk. 

Furthermore, earlier studies showed that a larger mean molecular weight 
($>$ 2.4, i.e. solar or super-solar metallicity) may be required for disk 
fragmentation to occur as this reduces the compressional heating 
({\it Mayer et al.}, 2007).  This study was carried out for the inner 
disk where small opacity changes are not as important (big changes are 
required for any effect to be seen, as illustrated above).  Therefore 
the effect of a reduced compressional heating was more important.  
However, varying both parameters for the outer disk, which may be the most interesting region for disk fragmentation, has not yet been conducted.
Variations of molecular weight from the
nominal solar value ($\sim 2.4$) to larger values ($2.6-2.7$) might occur as a result of the sublimation of ice coatings on grains as those grains pass through spiral shocks, if the dust-to-gas ratio is increased by at least an order of magnitude relative to mean nebular values ({\it Podolak et al.}, 2011). 
This increase could occur, e.g., through dust migration into arms via gas drag. 
Further investigation of the topic 
by coupling chemistry and hydrodynamical simulations will be a major task
of future research on {\it DI}. 

Furthermore, the effects of opacity in the spiral arms of self-gravitating 
disks also play an important role. Once fragments form, their survivability 
needs to be considered; it is affected by metallicity via a variety of 
processes that can determine changes in the opacity inside the collapsing 
clump and therefore affect its ability to cool and contract.

In summary, from a disk lifetime perspective, planet formation in low 
metallicity systems has to occur faster than those in high metallicity 
systems - making gravitational instability a more viable formation 
mechanism in such environments than core accretion.  However, a clear 
trend between the efficiency of planet formation via gravitational 
instability and metallicity has not yet been established since there are 
competing effects whose interplay at large radii (where gravitational 
instability is more likely to operate) is yet to be determined.  For the 
outer disk, opacity changes are clearly important and suggest lower 
metallicity favors fragmentation.  However its importance relative to 
the effects of the mean molecular weight and that of line cooling is unclear.  From the observational perspective the statistics 
are simply too low to rule in or rule out the possibility that gravitational 
instability is important at the low metallicity end ({\it Mortier et al.}, 2013).

\noindent
\subsubsection{\textbf{Clump Evolution}}\label{sec:clump_collapse}

The initial mass of the protoplanets that are formed
via gravitational instability is not well constrained but estimates
suggest a mass range from $\sim$ 1 M$_{\text{Jup}}$ to $\sim$ 10 M$_{\text{Jup}}$ 
({\it Boley et al.} 2010; {\it Forgan and Rice}, 2011; {\it Rogers and Wadsley}, 2012). This uncertainty
also leads to different predictions for the destiny
of such clumps. For example, while {\it Boley et al.}  (2010)
and {\it Cha and Nayakshin}  (2011) find that most clumps migrate
inward and get tidally destroyed at  several tens of
AU from the star, {\it Stamatellos and Whitworth} (2009) and {\it Kratter et al.} (2010b) find
that the formation of much more massive objects is possible
(brown-dwarfs of tens of M$_{\text{Jup}}$) at radial distances of
$\sim 100$ AU from the parent star.
Global disk simulations are therefore not yet able to predict the outcome of clump formation in a robust fashion. \par

After a gaseous clump has formed in the disk, its subsequent evolution--namely whether or not
a protoplanet with size, mass, density, and angular momentum comparable to a giant planet arises--depends on the initial conditions of the clump and on the nature of the 
formation process.
Clumps indeed start with a fairly well-constrained density and temperature profile, as well as with a distribution of angular momentum inherited from the disk material ({\it Boley
et al.} 2010). Their initial composition can vary from sub-nebular to 
highly-enriched compared to the star, depending on the birth environment 
in which the protoplanet has formed (see \S \ref{sec:predicted_composition} for details).  
The composition of the clump also affects the evolution since it determines
the internal opacity (see {\it Helled and Bodenheimer}, 2010). The clump should also become convective, fully or partially, depending on its temperature structure and composition.

Once formed, the evolution of clumps 
 consists of three main stages. 
During the first stage, the newly-formed planetary object is cold and extended with a 
radius of a few thousand to ten thousand times Jupiter's present 
radius (R$_{\text{Jup}}$), with H being in
 molecular form (H$_2$). The clump contracts quasi-statically on time scales of $10^4-10^6$ years,
depending on mass, and as it contracts, its
 internal temperature increases. When a central temperature of $\sim$ 2000 K is reached, the molecular H dissociates, and a dynamical collapse of the entire protoplanet occurs 
(the second stage). The extended phase is known as the {\it pre-collapse stage} ({\it DeCampli and Cameron},
 1979), and during that phase the object is at most risk of being destroyed by tidal disruption
 and disk interactions. After the dynamical collapse, the planet becomes compact and dense, with
 the radius being a few times R$_{\text{Jup}}$. During this third stage it is therefore less likely to 
be disrupted, although the planetary object still has the danger of falling into its parent star
 due to inward migration, as in the {\it CA} model. The protoplanet then continues to cool and contract on a much longer
 timescale ($10^9$ years). At this later stage, the evolution of the planet is similar for both {\it CA} and {\it DI}. \par 
 
Helled and collaborators ({\it Helled et al.~}2006; 2008; 
{\it Helled and Bodenheimer}, 2010; 2011) have studied the $1D$ evolution 
of spherically symmetric, non-rotating clumps assuming various masses and compositions. 
Essentially, similarly to {\it CA} models, a system of differential equations analogous to those of stellar evolution is solved.
These models were used to investigate the possibility of core formation and planetary enrichment, as we discuss below, and in addition, to explore the change in the pre-collapse evolution under different assumptions and formation environments.   
The pre-collapse timescale of protoplanets depends on the planetary mass, composition, and distance from the star, and it can vary from a few times
$10^3$ to nearly $10^6$ years. 
These timescales are long enough that the clump might be exposed to several harmful mechanisms while it is still a relatively low-density object, such as tidal mass loss during inward migration or photoevaporation by the central star and/or by neighboring massive stars as we discuss below. However, during this stage, physical processes such as grain settling and planetesimal capture, which can significantly change the planetary composition, can occur. 
Typically, massive planets 
evolve faster, and therefore have a higher chance to survive as they 
migrate towards the central star. For massive objects ($> 5~{\text M}_{\text {Jup}}$), the 
pre-collapse timescales can be reduced to a few thousand years if the protoplanets have
low opacity (and therefore more efficient cooling), due to either low metallicities or due to opacity reduction via grain growth and settling ({\it Helled and Bodenheimer}, 2011). Likewise the
pre-collapse timescale can be increased to 
more than $10^5$ year if the metallicity is as large as three times solar and the protoplanet has only a few Jupiter masses.

The pre-collapse evolution of protoplanets with masses of 3 and 10
${\text M}_{\text {Jup}}$ is presented in Fig.~4. As can be seen from the figure, more massive protoplanets have
 shorter pre-collapse time-scales. The initial radius increases  with increasing
 planetary mass, as more massive objects can be more extended and still be
 gravitationally bound due to stronger gravity. The central temperature however,
 is higher for more massive bodies. 

\begin{figure}[h!]
   \centering
    \includegraphics[width=7.1cm]{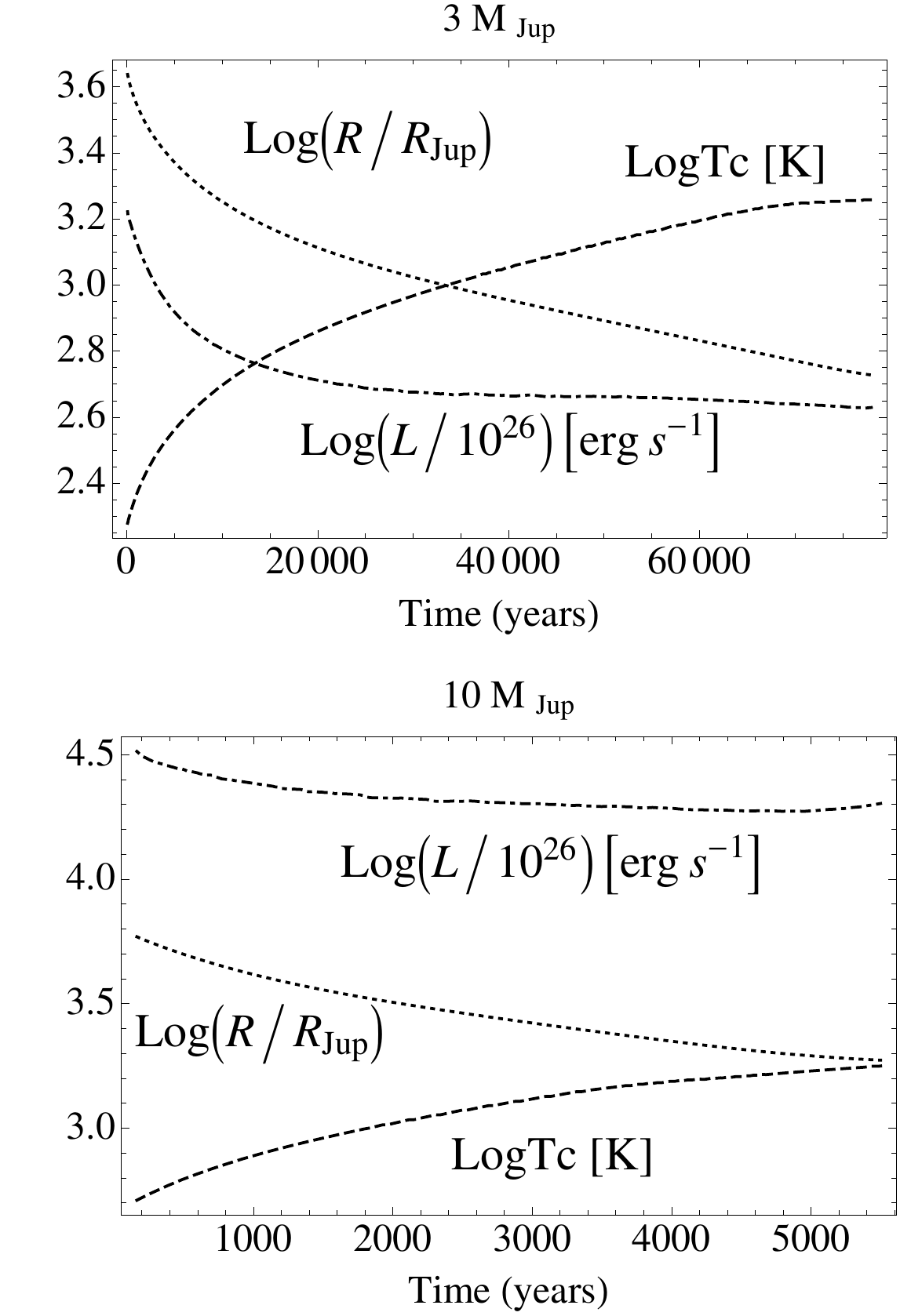}
    \caption[err]{
    {\small
    Pre-collapse evolution of 3 and 10 ${\text M}_{\text {Jup}}$ assuming solar composition and
Ê Ê Êinterstellar dust opacity. The luminosity, radius, and central 
temperature vs.~time are presented. Adapted from {\it Helled and Bodenheimer} (2010).
}
}
\end{figure}

3D hydrodynamical collapse simulations of clumps formed by {\it DI} are in their early stages. In this case clumps are
identified in simulations of global, self-gravitating protoplanetary disks, extracted from the
disk with a numerically careful procedure, and then
resampled at much higher resolution in order to follow the collapse to much higher
densities ({\it Galvagni et al.}, 2012).
Such clumps possess appreciable amounts of angular momentum with the highest values being above the expected threshold
for the bar instability ({\it Durisen et al.},~1986). Indeed a bar-like mode is observed to arise in the
clumps with very fast rotation, while in the others only strong spiral arms develop.

The current 3D simulations are purely hydrodynamical, with no account for solid-gas interactions
such as planetesimal accretion and dust sedimentation, at variance with what can be done
with the aforementioned 1D models. Only the solar metallicity case has been considered so far.
On the other hand, they follow, in the most general way possible, the hydrodynamics and self-gravity of a collapsing clump, and include also radiative
cooling and compressional/shock heating. Strong compression and shocks are not 
included by construction in 1D models. 
Radiative cooling is employed with different degrees of sophistication. 
The simulations adopt an EOS that takes into account the rotational, vibrational, and dissociation modes of H$_2$, and employ a simple cooling function ({\it Boley
et al.}, 2010), that interpolates
between optically thin and optically thick regimes. In the study by {\it Galvagni et al.} (2012) the pre-collapse timescales were found to be $\sim$1000 years, comparable to the shortest found in 1D calculations.

\subsubsection*{Opacity/Metallicity and the Contraction Of Protoplanets}
 The opacity plays an important role also in terms of cooling during
the post-formation contraction.  Thus, metallicity has a direct impact on the   
pre-collapse evolution of the newly-formed planets, and possibly on their
survival.
 {\it Helled and Bodenheimer} (2011) have shown that when the planetary 
opacity is scaled with the stellar metallicity, the pre-collapse timescale 
is proportional to planetary metallicity. Shorter pre-collapse stages of
 metal-poor protoplanets can support their survival.~Metal-rich protoplanets 
are more vulnerable to destruction; 
 however, if they do manage to survive, they have a better
 opportunity to accrete solid planetesimals and form heavy-element cores.  \par

When grain coagulation and sedimentation are included, it is found that the 
pre-collapse stages are 
 significantly shorter for all the planetary masses and 
metallicities considered. The time scale is found to be $\sim 10^3$  
years for masses between 3 and 7 M$_{\text{Jup}}$, and is relatively insensitive to 
planetary
 composition ({\it Helled and Bodenheimer}, 2011).  It is found that 
the pre-collapse evolution of  a metal-rich
 protoplanet can actually be shorter than that of a metal-poor planet, 
 a result of  very efficient
 opacity reduction caused by the larger amounts of grains initially present in the
 atmosphere, leading  to rapid grain growth and settling.  
The shorter time scales lead to two consequences: (1) a reduction in the
risk of clump disruption, and (2) smaller final masses, 
 because clumps accrete gas (and solids) most efficiently during their 
early evolution.  On the other hand, the short times would 
lead to less-significant enrichment with heavy elements, at least via planetesimal capture, and would
 suppress the formation of cores in these objects. 
Whether {\it DI} or {\it CA}, the outcome of planet formation seems to be driven by solids. \par

\noindent
\subsubsection*{The Effect of the Disk on the Planetary Evolution}

Recently, {\it Vazan and Helled} (2012) modeled the pre-collapse evolution of protoplanets coupled directly to
disk models and explicitly included accretion onto the protoplanet. The change in the pre-collapse timescale of a Jupiter-mass protoplanet for various radial distances was presented, and it was found that 1 M$_{\text{Jup}}$ protoplanet cannot evolve and contract at radial distances smaller than $\sim$11 AU due to the influence of the disk. This is due to the sufficiently high pressure and temperature of the disk, which prevent the protoplanet from contracting to reach a dynamical collapse and to become a gravitationally bound object.  While the exact radial distance for which contraction is no longer possible depends on the assumed disk model, the results of {\it Vazan and Helled} (2012) suggest that the presence of the disk has a non-negligible effect on the planetary evolution, and that the pre-collapse timescale of protoplanets can significantly lengthen for protoplanets as they get closer to the star. The pre-collapse evolution timescale of a 1 M$_{\text{Jup}}$ protoplanet, with solar composition and interstellar grain opacity, was found to range between 10$^5$ and $10^6$ years for radial distances between 50 and 11 AU, respectively. A Saturn-mass protoplanet was found to dissipate at a radial distance of $\sim$ 12 AU, while 3 M$_{\text{Jup}}$ and 5 M$_{\text{Jup}}$ protoplanets are found to  dissipate only at 9 AU, and 7 AU, respectively. Clearly, the influence/presence of the protoplanetary disk is less significant for more massive protoplanets. 

Since protoplanets are embedded in the gaseous disk, gas can be accreted as they evolve, therefore, the effect of gas accretion on the planetary evolution was investigated as well. It was shown that an increase of mass due to efficient gas accretion results in a faster contraction. It is also found that the planetary location has no significant influence on the planetary evolution when high accretion rates are considered. High accretion rates lead to significantly higher luminosity which lead to shorter pre-collapse timescales (see {\it Vazan and Helled}, 2012 for details). \par

The conclusion of  {\it Vazan and Helled} (2012) that the formation of
planets by {\it DI} is limited to relatively large
radial distances is in good agreement with previous research ({\it Rafikov}, 2007; {\it Boley}, 2009), but for the first time, it was based
on planetary evolution considerations. However, if protoplanets are
formed at large radial distances and migrate inward {\it after} the
dynamical collapse, when the protoplanets are denser and more compact,
the planets could survive at small radial distances. Alternatively,
protoplanets could form and survive at small radial distances if their
initial masses are sufficiently large ($>>$ 1 M$_{\text{Jup}}$). \par

\noindent
\subsubsection*{Clump Migration in Gravitationally Unstable Disks}

The angular momentum exchange and resulting migration of planets in
gravitationally unstable disks varies noticeably from migration in
laminar disks. Both {\it Baruteau et al.} (2011) and {\it Michael et al.} (2011) 
found, using independent methods, that planets in gravitationally
unstable disks can migrate rapidly inward on roughly Type I migration
timescales (a few orbital times). In addition to the
usual torques that are exerted on planets, stochastic torques are
present due to disk turbulence, which may kick a planet inwards or
outwards (the effects of such stochastic kicks are greater for lower
mass planets).  Furthermore, there are some indications that planets
may stall ({\it Michael et al.},  2011) or even open gaps in the
disk ({\it Zhu et al.}, 2012; {\it Vorobyov}, 2013), resulting in slower
migration; although, the opening of clear gaps is not efficient for
planets even as massive as 5 M$_{\rm Jup}$  due strong turbulence in the disk
resulting from gravitational instabilities and the stochastic orbital evolution of the planet
(e.g., {\it Baruteau et al.}, 2011).  Moreover, the evolution of the planet
itself can also affect the migration history.  For example,
mass-losing planets may suffer from an outward torque resulting in
outward migration ({\it Nayakshin and Lodato}, 2012).  Clearly, giant planet
formation models that include planetary evolution and migration
self-consistently are required to provide a more complete picture on
the formation and evolution of  {\it DI } planets.

\noindent
\subsubsection{\textbf{Composition of Protoplanets}}\label{sec:predicted_composition}

Planets formed by {\it DI} are often thought to have stellar abundances. Recent studies, however, show that the composition of {\it DI} planets can range from sub- to super- stellar.  

\subsubsection*{Metal enrichment from birth}

Spiral structure creates local gas pressure maxima that can collect solids 
through aerodynamic forces (e.g., {\it Weidenschilling}, 1977; {\it Rice et al.}, 2006; {\it Clarke and Lodato, 2009}), increasing 
the local solids-to-gas ratio. Spiral arms also act as local potential 
minima, which, when coupled with gas dissipation, can also help to maintain 
a high solid concentration.  Because clumps are born from the fragmentation 
of spiral arms,  they have the potential to form in regions of high solid 
concentrations.  Hydrodynamics simulations with gas-solid coupling show 
that fragments can indeed be born with super-stellar metallicities 
({\it Boley and Durisen}, 2010; {\it Boley et al.},~2011).  However, the 
degree of enrichment strongly depends on the size distribution of solids.  
Very small solids will be well-entrained with the gas, and will not show a 
high degree of concentration, while very large solids have long coupling 
timescales, and are not  expected to be captured by transient non-axisymmetric 
structure, such as spiral arms. The stopping time of a solid in a gaseous 
disk $t_s \sim \rho_s s/(\rho_g c_s)$ for a solid of size $s$ with 
internal density $\rho_s$ embedded in a gas with density $\rho_g$ and sound 
speed $c_s$.  Solids that are expected to have the greatest degree of 
trapping in spiral structure have stopping times that are within a factor of a few of their orbital period.  In a self-gravitating disk, maximum trapping corresponds to rock-size objects (10's of cm) and boulders ($\sim 100$ m). 

If most of the solids are in the 10 cm to 100 m size range, clumps can be 
enhanced at birth by factors of $\sim $1.5 - 2 ({\it Boley and Durisen}, 2010).
The overall enhancement is non-trivial, but fairly modest because some 
solid-depleted gas becomes mixed with the spiral arm during the 
formation of the clump. On the other hand, if most of the solids are in 
km-size objects or larger, then depletion at birth may actually be a 
possibility, as the solids do not necessarily follow the contraction of gas.  
The type of object that forms initially depends, in part, on the evolution of solids. 

\subsubsection*{Metal enrichment via planetesimal capture}

Another mechanism in which {\it DI} planets can 
be enriched with heavy elements is planetesimal accretion. Clumps that form 
in the disk are surrounded by planetesimals which are affected by the 
gravitational field of the protoplanet and can therefore capture a large 
portion of solid material.~{\it Helled et al.} (2006) and 
{\it Helled and Schubert} (2009) have shown that the final composition 
of the protoplanet (clump) can vary substantially when planetesimal capture is considered.  
The available solid mass for capture depends on the solid surface density 
$\sigma_s$ at the planetary location. This density changes with the disk mass, 
its radial density profile, and stellar metallicity.
The total available mass of solids in the planet's feeding zone depends strongly on $a$, on the radially decreasing $\sigma_s(a)$,
and weakly on the planetary mass. The actual planetary enrichment depends on the ability of the protoplanet to capture 
these solids during its pre-collapse contraction. The final mass that can be captured also depends on the physical properties of the planetesimals such as their sizes, compositions, and relative velocities. 

The planetesimal accretion rate is given by equation (\ref{eq:accrete}), but here the
capture radius is significantly larger. Enrichment by planetesimal capture 
is efficient as long as the protoplanet is extended (a few tenths of an AU). 
It then fills most of its feeding zone, so planetesimals are slowed 
down by  gas drag, and are absorbed by the protoplanet. Therefore the 
longer the clump remains in the pre-collapse stage, the longer the clump will be able to accrete. 
{\it Helled et al.} (2006) have shown that a 1 M$_{\text{Jup}}$ protoplanet at 
5.2 AU with $\sigma_s=10$ g cm$^{-2}$ could accrete up to 40 M$_{\oplus}$ 
of heavy elements during its pre-collapse evolution.   
{\it Helled and Schubert} (2009) have shown that a 1 M$_{\text{Jup}}$  protoplanet 
formed between 5 and 30 AU can accrete 1--110 M$_{\oplus}$ of 
heavy elements, depending on disk properties, and concluded that the final 
composition of a giant planet formed by {\it DI} is strongly 
determined by its formation environment. 

Enrichment of massive protoplanets at large radial distances such as those discovered around HR 8799 ({\it Marois  et al.}, 2008), where planets are likely to form by {\it DI}, has been 
investigated by {\it Helled and Bodenheimer} (2010). Since the timescale 
of the pre-collapse stage is inversely proportional to the  square of the
planetary mass, massive protoplanets would have less time to accrete solids. 

Clearly, the final composition of protoplanets can change considerably 
depending on the 'birth environment' in which the planet forms. The large variations in heavy 
element enrichment that are derived under different assumed physical conditions could, in 
principle, explain the diversity in the derived compositions of 
gas giant exoplanets. Since higher surface density leads to larger 
heavy element enrichments, the {\it DI} scenario is also  
consistent with the correlation between heavy element enrichment and stellar 
metallicity ({\it Guillot et al.}, 2006; {\it Burrows et al.}, 2007), 
highlighting the {\it danger in assuming that metallicity correlations are 
 prima facie evidence in support of the core accretion model}. 
 Note, however, that the correlation between the {\it probability of 
detection} of a giant planet and the metallicity of the star is a separate
issue.

\subsubsection*{Enrichment through tidal stripping}

Another mechanism which can lead to planetary enrichment in the {\it DI} scenario is enrichment through tidal stripping, 
which is one outcome of the tidal disruption/downsizing hypothesis 
({\it Boley et al.}~2010; 2011; {\it Nayakshin}, 2010).  Clumps create 
significant potential wells that allow solids to settle toward their 
center. This settling can lead to the formation of heavy-element cores 
(see \S \ref{sec:core_formation}), which will deplete the outer 
envelope of the clump.   If the clump can be stripped of this metal-poor 
gas, the final object can become significantly enhanced in metals, 
producing objects that have extremely massive cores for their bulk mass 
(e.g, HAT-P-22b). These Brobdingnags may be a natural outcome of the {\it DI} model for the following reasons.  (1) Clumps that form through 
fragmentation roughly fill their Hill sphere, to a factor of a few.  At 
distances of $\sim 100$ AU, this implies initial clump sizes $\sim 5$ AU.  
Clumps are highly susceptible to migration through the disk via clump-clump 
scattering ({\it Boley et al.}~2010) and clump-disk torques 
({\it Baruteau et al.}~2011; {\it Michael et al.}~2011).  If the inward 
migration rate is faster than the contraction rate, 
the clump can overfill its Hill sphere, resulting in significant mass loss.  If solid settling has occurred by this 
time, this overflow will preferentially remove  metal-poor gas.
\par

The various mechanisms for planetary enrichment in the {\it DI} 
model are  much more efficient when the planets are formed in a metal-rich 
environment, i.e., around metal-rich stars. 

\subsubsection{\textbf{Core Formation}}\label{sec:core_formation}

Protoplanets formed by {\it DI} can still have cores in 
their centers. However, unlike in the {\it CA} model, the 
existence of a core is not directly linked to the formation mechanism, and 
the core can also form {\it after} the formation of the planet. One way 
in which cores can form 
is via {\it enrichment from birth}. {\it Boley and Durisen} (2010) and 
{\it Boley et al.} (2011) have shown (building on the work of {\it Rice et al.}, 2004; 2006) that the solids which are collected 
by the forming planets tend to concentrate near the center, and can therefore 
end up as heavy-element cores. 

Another way in which cores can be formed is via grain coagulation and settling.
During the pre-collapse evolution the internal densities and 
temperature are low enough to allow grains to coagulate and sediment to 
the center
({\it DeCampli and Cameron}, 1979; {\it Boss}, 1997). 
Both of these studies, however, assumed that the planetary interior is 
radiative while now protoplanets are found to be convective 
({\it Wuchterl et al.}, 2000; {\it Helled et al.}, 2006). 

A detailed analysis of coagulation and sedimentation of 
grains of various sizes and compositions in evolving protoplanets including the presence of convection have been presented by 
{\it Helled et al.} (2008) and {\it Helled and Schubert} (2008). 
It was confirmed that silicate grains can grow and sediment to form a core both for 
convective and non-convective envelopes, although the sedimentation times
of grains with sizes smaller than about 1 cm are substantially longer if the 
envelope is convective. The reason is that grains must grow and 
decouple from the convective flux in convective regions in order to 
sediment to the center. Small icy/organic grains were found to dissolve 
in the planetary envelope, not contributing to core formation. Settling 
of volatiles into the planetary center is possible only if their sizes 
are sufficiently large (meters and larger). 
\par

The settling timescale for solids in a given clump can be roughly estimated 
by assuming that solids will move inward at their terminal speed, which is 
roughly $g t_s$, where $g$ is the gravitational acceleration.  Consider 
a 3 M$_{\text{Jup}}$ clump with a size of $\sim 3$ AU. The corresponding pre-collapse evolution is of the order of $10^4$ years. 
These values are meant only 
to highlight one of many possible configurations.  The average density 
for this structure is $\bar{\rho}\sim 2\times10^{-11}$ g cm$^{-3}$, with average 
temperature $\bar{T}\sim m_p \mu GM/(k R ) \sim 250$ K, where $m_p$ 
and $\mu$ are the proton mass and the mean molecular weight (2.33 here), 
respectively.  Using these values, the radial settling speed is, to within 
factors of order unity, $v_r \sim -[\rho_s s/(\rho_g c_a)] GM/R^2$. 
For our envisaged clump, $v_r\sim7\times10^{-4}$ AU/yr for $s\sim 1$ cm 
and $\rho_s\sim 3$ g cm$^{-3}$. Settling can take place within a few $10^3$ years, in principle, before 
a clump collapses due to H$_2$ dissociation, opening up the possibility 
of enriching a clump through tidal stripping of metal-poor gas. Understanding 
how convection  and dust evolution alter this simple picture requires detailed simulations.
\par

A third mechanism for core formation is settling of larger bodies, i.e.~planetesimals that are captured by the protoplanet. 
Core formation is favorable in low-mass protoplanets due to their longer contraction timescales which allows more planetesimal accretion. In addition, they have lower internal temperatures, and  lower convective velocities which support core formation ({\it Helled and Schubert}, 2008). The final core mass changes with the 
planetary mass, the accretion rate of solids (planetesimals), and radial 
distance. If planetesimals are captured during the pre-collapse stage, 
they are likely to settle to the center, and therefore increase the core 
mass significantly. As a result, there is no simple prediction for the core 
masses of giant planets in the {\it DI} model, however simulations suggest a possible range of zero to hundred M$_{\oplus}$.  
In addition, it seems that the cores are primarily  
composed of refractory materials with small (or no) fractions of 
volatile materials (ices/organics). If grain settling works together with, e.g., enrichment at birth, then it may be possible to form much larger cores than what one would expect from pure settling.   \par 

\subsubsection{\textbf{Mass Loss}}

Protoplanets formed by {\it DI} could 
lose substantial envelope mass, which can lead not just to enrichment through 
removing dust-depleted gas, but might also lead to formation of rocky 
cores/planets ({\it Nayakshin}, 2010; {\it Boley et al.},~2010).  
The cooling and migration timescales determine the degree of mass loss 
that can occur and where in the disk mass loss can 
commence. 

The planet-star separation at which the mass loss commences depends on the mass and the age of the protoplanet, as well as the conditions in the protoplanetary disk itself. If H is still predominantly molecular in the protoplanet (very young and/or relatively less massive, e.g., $M_p\simlt$ 5 M$_{\text{Jup}}$ planet), then disruption is likely to occur at distances of $\sim 1$ AU, although the survival of clumps at this evolutionary stage {\it and} this distance has been shown to be unlikely ({\it Vazan and Helled}, 2012). If the planet is additionally puffed up by the internal energy release such as a massive core formation ({\it Nayakshin and Cha}, 2012) or by external irradiation ({\it Boss et al.},~2002) then tidal disruption may occur at much greater distances, e.g., tens of AU where clumps are more likely to exist.

On the other hand, if the protoplanet is far along in its contraction 
sequence (more massive $M_p\simgt$ 10 M$_{\text {Jup}}$ and/or older protoplanets), then 
it may be in a ''second stage'', where H is atomic and partially ionized. 
In this case the protoplanet is at least an order of magnitude more compact 
and cannot be tidally disrupted unless it migrates as close as 
$\sim 0.1$ AU to its star ({\it Nayakshin}, 2011). It has been suggested that such tidal disruption 
events may potentially explain the population of close-in planets, both terrestrial and giant (see{\it Nayakshin}, 2011 and references therein).

Because fragments become convective as they contract 
their internal structure can be approximated by a polytrope with $n=2.3$, which is appropriate for a mixture of hydrogen and helium. 
In this regime, molecular protoplanets should have runaway disruptions, 
leaving behind only the much denser parts, e.g., the cores. On the 
other hand, tidal disruptions of protoplanets in the immediate vicinity 
of the protostar, e.g., at $\simlt 0.1$ AU, are much more likely to be 
a steady-state process.  {\it Nayakshin and Lodato} (2012) suggest that 
the protoplanets actually migrate outward during this process. The accretion 
rates onto the protostars during a tidal disruption process are as large as 
$\sim 10^{-4}~ \msun$~yr$^{-1}$, sufficiently large to present a natural 
model to explain the FU Ori outbursts of young protostars ({\it Hartmann and 
Kenyon}, 1996; {\it Vorobyov and Basu}, 2006; {\it Boley et al.}, 2010).
\par 

\bigskip
\noindent
\section{\textbf{POST-FORMATION EVOLUTION}}
\bigskip

The formation phase of a giant planet lasts for $\sim10^4-10^6$ years, depending on the formation model, 
while the following phase of slow contraction and cooling takes place over
billions of years. Thus it is the latter phase in which an exoplanet is most likely to be detected. The salient features are the gradual
decline in luminosity with time, and the mass-luminosity relation which
gives luminosity of a planet roughly proportional to the 
square of the mass, so the prime targets for direct detection
are relatively massive  planets around  the younger stars.

After accretion is terminated, the protoplanet evolves at constant mass  on a time
scale of several Gyr with
energy sources that include  gravitational contraction, cooling of the warm
interior, and  surface heating from the star. The latter source is important
for giant planets close to their stars because the heating of the surface layers
delays the release of energy from  the interior and results in a somewhat larger
radius at late times than for an unheated planet.
For giant planets close to their stars, tidal dissipation in the planet, caused
by circularization of its orbit and synchronization of its rotation with its
orbital motion, can provide a small additional energy source ({\it Bodenheimer et al.}, 2001). In addition, magnetic field generation in close-in planets can result in Ohmic dissipation in the interior leading to inflation ({\it Batygin and Stevenson}, 2010). 

The initial conditions for the  final phase depend only weakly on
the formation process. In the 
{\it DI} model the evolution goes through a phase of
gravitational collapse induced by molecular dissociation, and equilibrium is
regained, 10$^5$ to 10$^6$ yr after formation, at a radius
of only 1.3 R$_\mathrm{Jup}$ for the case of 1.0 M$_\mathrm{Jup}$ 
({\it Bodenheimer et al.}, 1980).
In the case of {\it CA} models, calculations ({\it Bodenheimer 
and Pollack}, 1986; {\it Bodenheimer et al.}, 2000) show
that  after the formation phase, which lasts a few Myr, the radius declines to $\approx 10^{10}$ cm, or 1.4 R$_\mathrm{Jup}$, 
on a time scale of 10$^5$ yr for 1 M$_\mathrm{Jup}$. This comparison has not been made, however, forÊ
planets in the 10 M$_\mathrm{Jup}$ range.

An important point is that in both cases there is an accretion shock on
the surface of the planet. In the case of {\it DI} it 
occurs during the collapse induced by molecular dissociation. The outer 
layers accrete supersonically onto an inner hydrostatic core. 
In the case of {\it CA} it results from accretion of disk material
onto the protoplanet during the phase of rapid gas accretion when the
planetary radius is well inside the effective radius.
The heat generated just behind the shock escapes the object by radiation
through the infalling envelope. Thus much of the gravitational energy
liberated by the infalling gas does not end up as internal energy of
the planet: it is lost. The planet starts the final phase of
evolution in a stage of relatively low entropy, sometimes referred to
as a \emph{cold start}. 
The actual entropy of the newly-formed planet is
somewhat uncertain, as it depends on the details of how much energy is
actually radiated from behind the shock and how much is retained in
the interior of the planet.

However many numerical simulations do not take into account the formation
phase in the calculation of the final contraction phase. 
They simply assume that the evolution starts at a radius 2--3 times that of
the present Jupiter, and assume the model is fully convective at that
radius. The shock dissipation and radiation of energy is not taken into
account, and the result is that the initial  models are warmer and have
higher entropy than the `cold start' models. They are known as
`hot start' models. After a certain amount of time, both types of models
converge to the same track of radius or luminosity versus time, but
at young ages the results under the two assumptions can be quite different,
depending on the mass of the planet. Jupiter-mass planets
converge to the same track after 20 Myr; however 10 M$_\mathrm{Jup}$ planets 
take
over $10^9$ yr to converge ({\it Marley et al.}, 2007). Thus young, relatively massive planets
calculated according to the `cold start' are considerably fainter than
those, at the same age, calculated with the `hot start'. \par 

Note that it is often assumed that, as a result of gas accretion through a shock, planets formed by 
{\it CA} finish their formation with lower luminosities than 
 planets formed via {\it DI}. This conclusion, however, depends (i) on the efficiency of energy radiation through the shock, and
    (ii) on the core mass of the planet at the end of accretion. {\it Mordasini et al.} (2012b) show that a {\it CA} calculation
    in which no energy is radiated from the shock produces a model
    very close to a 'hot start'. Furthermore, 'cold start'
    calculations show that the entropy and luminosity of a planet
    after formation depend on its core mass ({\it Bodenheimer et al.}, 2013).
    For example, a 12 M$_\mathrm{Jup}$ planet (which is in the
   suspected mass range for directly imaged exoplanets), just after
    formation, could have a luminosity ranging from log $L$/L$_\odot= -4.8$
    to -6.6, corresponding to core masses ranging from 31 to 4.8 M$_\oplus$.
   For lower final masses or higher core masses, the luminosities can be even higher ({\it Mordasini}, 2013).
    In fact, both formation models could result in either "cold" or
    "hot" starts. 
In order to better understand the evolutions of giant planets when accretion and shock calculations are included the development of 3D hydrodynamical models is required. 
\par

Future planetary evolution models should include physical processes such as core erosion, rotation, settling, etc., which are typically not included in evolution models due to their complexity. Such processes could change the contraction histories of the planets as well as the final internal structure. 
While we are aware of the challenge in including such processes in planetary evolution models, we hope that by the next  {\it Protostars and Protoplanets} chapter on giant planets, progress in that direction will be reported. 

\bigskip
\section{\textbf{SUMMARY OF GIANT PLANET FORMATION MODELS}}
  
\subsection{\textbf{Core Accretion}}
 A giant planet forming according to the {\it CA} model
goes through the  following steps. 
\begin{itemize}
\item Accretion of dust particles and planetesimals results in a solid core
of a few M$_\oplus$, accompanied by a very low mass gaseous envelope.
\vspace{-2.mm}
\item Further accretion of gas and solids results in the
mass of the envelope increasing faster than that of the core until a
crossover mass is reached.
\vspace{-2.mm}
\item Runaway gas accretion occurs with relatively little
accretion of solids.
\vspace{-2.mm}
\item Accretion is terminated by tidal truncation (gap formation)
 or dissipation of the nebula.
 \vspace{-2.mm}
\item  The planet contracts and cools at constant mass to its  present state.
\end{itemize}

The strengths of this scenario include
\begin{enumerate}
\vspace{-2.mm}
\item The model can lead to both the formation of giant and icy planets. 
\vspace{-2.mm}
\item The model can explain the correlation between stellar metallicity and giant planet occurrence {\it and} the correlation between the stellar and planetary metallicity. 
\vspace{-2.mm}
\item The model is consistent with the enhancement of the heavy
element abundances in Jupiter, Saturn, Uranus and Neptune. 
\vspace{-2.mm}
\item The model predicts that giant planets should be rare around
stars with lower mass than the Sun. 
There are in fact some giant planets around low-mass stars, but the frequency
(for masses above 0.5 M$_{\text{Jup}}$) is only about 3\% for 
stars of about 0.5 M$_\odot$ ({\it Johnson et al.}, 2010), at solar
metallicity.
\vspace{-2.mm}
\item The long formation time, caused by the slow gas accretion during
 Phase 2 can be considerably reduced when the effect
 of grain settling and coagulation on the dust opacity is taken
into account ({\it Movshovitz et al.}, 2010). Similarly, including migration resulting  from
disk-planet angular momentum exchange can suppress phase 2, therefore decreasing
substantially the formation time. 
\end{enumerate}

A number of problems also emerge:

\begin{enumerate}
\vspace{-2.mm}
\item The model depends on the processes of planetesimal formation and early embryo growth which are poorly understood. 
\vspace{-2.5mm}
\item The time scale for orbital decay of the planet due to
type I migration, although uncertain, may still be
shorter than the time required to build up the core to
several M$_\oplus$.   Further details on these processes are described in
the chapter by {\it Baruteau et al.} 
\vspace{-2.5mm}
\item The model faces extreme difficulties in explaining planets
around stars of very low heavy-element abundance or  massive giant
planets at radial distances  greater than about 20 AU.
\vspace{-2.mm}
\item In the absence of migration the formation time scale
is close to the limits imposed by observations of protoplanetary disks.  
\vspace{-2.mm}
\item Final envelope abundances and gas giant formation time scales are dependent on the dynamics of planetesimals which is still very uncertain. 
\vspace{-2.mm}
\item The opacity of the envelope plays a critical role in gas giant planet formation time scales. The actual opacity of accreted gas onto
a growing gas giant and the role grain growth plays in the evolution of solids in the envelope still have significant uncertainties.
\end{enumerate}

\subsection{\textbf{Disk Instability}}

While as old as the {\it CA} paradigm, if not older, 
{\it DI} is taking considerably longer to mature as a theory.  
In part this is due to the complexities of understanding the conditions under 
which disks can fragment, but is also in part due to 
the realization that additional branches of the theory are required,
including recognizing possible paths for forming rocky objects.

Generally speaking, the main weakness of the {\it DI} model is the uncertainty regarding whether clumps can indeed form in realistic disks, and even 
if they do form, it is not clear if they can actually survive and evolve to 
become gravitationally bound planets.  Giant planet formation by {\it DI} is conditioned by:

\begin{itemize}
\vspace{-2.mm}
\item Disks must at some point become strongly self-gravitating.  This can 
be achieved by mass loading sections of a disk through accretion of the protostellar envelope or through variations of the accretion rate in the disk itself.   
\vspace{-2.mm}
\item There must be significant cooling in the disk to increase the likelihood of fragmentation of spiral structure, although the exact limits are still being explored.
\end{itemize}

The main advantages to this mechanism are:

\begin{enumerate}
\vspace{-2.mm}
\item Fragmentation can in principle lead to a variety of outcomes, including gas giants with and without cores, metal rich and metal poor gas giants, brown dwarfs, and possibly terrestrial planets. These various outcomes depend on a complex competition between the cooling, dust coagulation and settling, accretion, and disk migration time scales. 
\vspace{-2.mm}
\item Planet formation can begin during the earliest stages of disk evolution, well within the embedded phase. 
\vspace{-2.mm}
\item {\it DI} may lead to many failed attempts at planet formation through fragmentation followed by tidal disruption, which could be linked with outburst activities of young protoplanetary disks, as well as to the processing of some of the first disk solids. 
\vspace{-2.mm}
\item {\it DI} can take place at large disk radii and in low metal 
environments, consistent with  the HR 8799 ({Marois et al.}, 2008) 
planetary system.
\end{enumerate}

The main disadvantages of this mechanism are:

\begin{enumerate}
\vspace{-2.mm}
\item The exact conditions that can lead to disk fragmentation and the frequency of real disks that become self-gravitating are not  well understood. 
\vspace{-2.mm}
\item Even if protoplanets can be formed by {\it DI}, whether they can survive (both tidal disruption and rapid inward migration) to become gravitationally bound planets is still questionable. 
\vspace{-2.5mm}
\item The model cannot easily explain the formation of intermediate-mass planets.
\vspace{-2.mm}
\item The model does not provide a natural explanation for the correlation of giant planet occurrence and stellar
     metallicity. 
\vspace{-2.mm}
\item Grain evolution is a principal branching mechanism for the different types of objects that form by {\it DI}. Thus, a predictive model for {\it DI} cannot be developed without a more complete model of grain physics. 
\vspace{-2.mm}
\item The detailed effects of mass accretion onto clumps and mass removal by tides must still be explored.  
\vspace{-2.mm}
\item As summarized above for {\it CA}, the detailed interactions between clumps and planetesimals, if present, must be further developed. 
\end{enumerate}

\subsection{\textbf{Core Accretion and Disk Instability: Complementarity}}

{\it CA} and {\it DI} are not necessarily competing processes.  
{\it DI} is likely most common during the early embedded phases of disk evolution ($\sim$ few $10^5$ yr), 
while {\it CA} occurs at later stages ($\sim$ few Myr).  
In this view, {\it DI} could represent the first trials of planet formation, which may or may not be successful.  If successful, it does not preclude formation by {\it CA} at later stages. 

{\it CA} may be the dominant mechanism for forming ice giants and 
low-mass gas giants, while {\it DI} may become significant for 
the high end of the mass distribution  of giant planets.  Nevertheless, it seems possible that both scenarios can lead to the formation of an object on either end of the mass spectrum.  

Overall, {\it CA} explains many of the properties of  Solar System planets and exoplanets.  Nevertheless, a number of exoplanets cannot naturally be explained by the {\it CA} model such as giant planets at very large radii and planets around metal-poor stars, both of which have been observed.
Models of both {\it CA} and {\it DI} are continuously rejuvenating, and we hope by {\it PPVII} will have a better understanding of the types and relative frequency of planets under those two paradigms.

\bigskip
\noindent
\section{\textbf{GIANT PLANET INTERIORS}}
\noindent
\subsection{\textbf{The Giant Planets in the Solar System}}

Although there are now hundreds of known giant exoplanets, we
can usually measure only the most basic parameters: their mass and/or radius.  While the mean density is a powerful diagnostic, it is insufficient to give a detailed picture of the planetary interior.  
At present, only the planets in our own Solar System can be studied in sufficient detail to allow us to form such a picture.  Details of these modeling efforts are presented in the chapter by Baraffe et al.~but because the results of such models are of great importance in helping us constrain formation scenarios, we devote this section to looking at some of the difficulties and limitations of interior models.

Models of the planetary interior have three main components that can be characterized as {\it simple physics}, {\it complex physics}, and {\it philosophy}.  In addition, the issues that need to be considered in each of these components will be different for the class of bodies that are mostly H-He, like Jupiter and Saturn, than for the class comprising Uranus and Neptune.  The latter contain substantial amounts of H-He as well, but the bulk of their mass is high-Z material.

\noindent
\subsubsection{\textbf{The Gas Giants: Jupiter \& Saturn}}

\subsubsection*{Simple Physics}
The basic components of any model are the conservation of mass and of momentum.  The former is written as
\begin{equation}\label{mcon}
\frac{dM}{dr}=4\pi r^2\rho(r)
\end{equation}
where $M(r)$ is the mass contained in a sphere of radius $r$ and $\rho(r)$ is the density at $r$.  Conservation of momentum is simply an expression of force balance, and is given by
\begin{equation}\label{pcon}
\frac{dP}{dr}=-\frac{GM(r)\rho(r)}{r^2}+\frac{2}{3}\omega^2r\rho(r)
\end{equation}
where $P$ is the pressure, and $\omega$ is the rotation rate.  The second term on the RHS is an average centrifugal term that is added to allow for the effect of rotation.  For Jupiter this amounts to a correction of $\sim$2\% in the interior, rising to $\sim$5\% near the surface.  For Saturn it is about twice that.  For Uranus and Neptune it is less than 1\% except near the surface.  As the inputs to the models become more precise, eqs.\,(\ref{mcon}) and (\ref{pcon}) must be replaced by their 2-D versions.

\subsubsection*{Complex Physics}
A description of how the temperature changes within the planet must be added to the framework described above, and this, in turn, depends on the mechanism of energy transport.  {\it Hubbard} (1968) showed that in order to support the high thermal emission observed for Jupiter, most of its interior must be convective, and therefore this region can be well represented by an adiabat.  It is generally assumed that the same is true for Saturn.  Although {\it Guillot et al.} (1994) have suggested the possibility that Jupiter and Saturn might have radiative windows at temperatures of $\sim$2000\,K, and {\it Leconte and Chabrier} (2012) have studied how composition gradients can modify the usual adiabatic  gradient through double diffusive convection, the assumption of a strictly adiabatic temperature gradient is still quite standard.

Even with the assumption of an adiabatic temperature gradient, there is still the non-trivial problem of computing the density profile.  Equations of state have been measured experimentally for H, He, H$_2$O, and some additional materials of interest, but there are still large uncertainties in the experiments themselves and in their theoretical interpretation (e.g., {\it Saumon et al.}, 1995; {\it Nettelmann et al.}, 2008; {\it Militzer et al.}, 2008). This leads to significant uncertainties in, for example, the estimated core mass of Jupiter and Saturn. The lack of knowledge on the depth of differential rotation in Jupiter and Saturn also introduces a major source of uncertainty to structure models. 

Another feature that is difficult to model is the effect of the magnetic field.  The effects in Jupiter and Saturn are expected to be small, and have mostly been ignored, but as measurements improve, including magnetic effects will become necessary.  These will include pressure effects that must be added to eq.\,(\ref{pcon}), as well as contributions to the viscosity that will affect the convective motions, and, as a result, the temperature gradient.  These motions may also contribute to differential rotation in these planets, and may themselves be driven by differential rotation. In return, magnetic field modeling may eventually allow us to set useful limits on location of conductive regions of the planet's interior (e.g., {\it Stanley and Bloxham}, 2006; {\it Redmer et al.}, 2011), and on the extent of the differential rotation. 

\subsubsection*{Philosophy}
Perhaps the most difficult aspect of modeling is the choice of overall structure and composition.  Traditional models of Jupiter and Saturn assumed a core of heavy material surrounded by a H-He envelope with some additional high-Z component ({\it Podolak and Cameron}, 1974; {\it Saumon and Guillot}, 2004), but more recent models have divided the envelope into a heavy element rich layer under a layer with more nearly solar composition ({\it Chabrier et al.}, 1992; {\it Guillot}, 1999; {\it Nettelmann et al.}, 2012).  The motivation for dividing the envelope is based on the idea that at high pressures not only does H change from the molecular to the metallic phase, but He may become immiscible in H and rain out to the deeper interior ({\it Stevenson}, 1975; {\it Wilson and Militzer}, 2010). The depletion of He in Saturn's upper atmosphere has been observed ({\it Gautier et al.}, 2006) and He rain has been advanced as an explanation for Saturn's high thermal emission ({\it Fortney and Nettelmann}, 2010).

The assumption of a core was originally based on the need to match the low gravitational quadrupole moments of these planets ({\it Demarcus}, 1958).  This assumption was later strengthened by the {\it CA} model, which requires a large heavy element core to initiate the capture of a H-He envelope.  However, for both Jupiter and Saturn, interior models without cores can be found. This highlights the fact that the model results are due in part to the underlying picture used by the modelers in deriving their results.  This has led to a second general approach to interior modeling where some function is assumed for the density distribution as a function of radius, and the free parameters of this function are chosen to fit the observables such as the average density, and the moments of the gravitational field.  One then tries to interpret this density distribution in terms of composition using equations of state. 

\subsubsection*{Ambiguities in the Models}
The reason the core mass varies so widely can be understood by considering a simplified {\it toy model} in which the density distribution is given by 
\begin{eqnarray}
\rho(r) =\rho_c~~~~r\leq R_c \nonumber \\
\rho(r)=\rho_0\left[1-\left(\frac{r}{R}\right)^{\alpha}\right] r>R_c
\end{eqnarray}
where $R_c$ is the core radius, $R$ is the planet's radius, and $\rho_c$, $\rho_0$, and $\alpha$ are constants.  $\alpha=2$ gives a fair representation of Jupiter's density distribution as derived from detailed interior models.  Once the core density $\rho_c$ is chosen, $R_c$ and $\rho_0$ are found by requiring that the distribution reproduce Jupiter's mass and inertia factor which are determined using the Radau approximation ({\it Podolak and Helled}, 2012).

\begin{figure}[ht]
\centerline{\includegraphics[width=8.4cm]{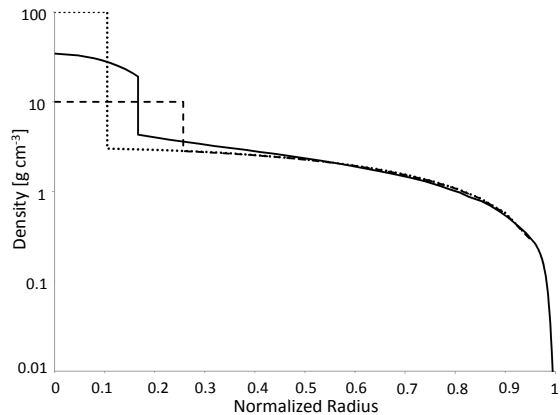}}
\caption{
{\small
Density vs.~radius for a realistic Jupiter model (solid curve), a toy model with $\rho_c$=10\,g\,cm$^{-3}$ (dashed curve) and a toy model with $\rho_c$=100\,g\,cm$^{-3}$ (dotted curve).  Although the core densities differ greatly, the densities for $r\gtrsim 0.5$ are nearly identical.}
}
\label{toy}
\end{figure}

As can be seen from Fig.~\,\ref{toy}, $\rho_c=10$\,g\,cm$^{-3}$ (dashed) or $\rho_c=100$\,g\,cm$^{-3}$ (dotted) give nearly identical values for the envelope's density, and this density distribution is very similar to one derived using realistic equations of state and matching the observed gravitational moments (solid).  In other words, not only are the mass and moment of inertia insufficient to distinguish between a core density of 10\,g\,cm$^{-3}$ and one of 100\,g\,cm$^{-3}$ and core masses differing by a factor of 1.5, but probing the density distribution in the outer parts of the planet is not enough either.  

We cannot measure the internal density distribution of a planet directly, but we can measure some of the moments of the distribution.  Because of the cylindrical hemispheric symmetries of a rotating planet the moments are even functions of only the radius $r$ and colatitude $\theta$ and are given by 
\begin{equation}\label{jn}
J_{2\ell}=-\frac{1}{Ma^{2\ell}}\int r'^{2\ell}P_{2\ell}(\cos\theta')\rho(r',\theta')d^3r'
\end{equation}
where $M$ is the mass, $a$ is the average radius, and $P_n$ is the $n$th-order Legendre polynomial.~ Interior models are constructed to fit the mass (essentially $J_0$) and as many of the $J_{2\ell}$'s as have been measured.  Although each higher order $J$ gives additional information on $\rho(r)$, this information is more and more strongly weighted by high orders of $r$.  
In fact, only $J_0$ (i.e. the mass) probes the region of the core.  The higher $J$'s are only sensitive to the density distribution in the envelope.  Since this is nearly independent of the core density, small changes in the envelope density caused by differences in the equation of state or in the assumed composition, can cause large changes in the calculated mass of the core which ranges between zero and up to $\sim$ 15 M$_{\oplus}$.

The composition of the envelope must also be treated with caution.  The density of a mixture of materials is given, to an excellent approximation, by the additive volume assumption.  If $\rho_i$ is the density of the $i$th component and $X_i$ is its mass fraction, then assuming the volume of the mixture is the sum of the volumes of the individual components gives
$1/\rho=\sum_i X_i/ \rho_i$. 
For Jupiter, H and He account for $\sim$90\% of the envelope by mass, while the high-Z component makes up the remainder.  On the other hand, the density of the high-Z component is much larger than that of H.  Thus the major contribution to $\rho$ is due to H.  As a result, if we try to match the density at some point in the envelope with a particular mixture, a small error in the equation of state of H may cause a much larger error in the estimated abundance of high-Z material ({\it Podolak and Hubbard}, 1998).

Finally, there is the ambiguity due to the uncertainty in the composition itself.  The same pressure-density relation can be well reproduced by different mixtures, even over a range of pressures.  Measurements of the magnetic field, which gives information about the conductivity as a function of depth may help break this degeneracy, but such measurements require detailed dynamo models for their interpretation, and the construction of such models is still in the early stages.  In addition, a more detailed knowledge of the planet's differential rotation (if any) is required.

\noindent
\subsubsection{\textbf{The Icy Giants: Uranus \& Neptune}}

Typically, giant planet formation models concentrate on Jupiter-like planets, while lower-mass planets, such as Uranus and Neptune, are simply considered to be "failed giant planets". These planets are 
more naturally explained by {\it CA}, in which the forming planets remain gas-poor due to their slow growth at farther distances where runaway gas accretion cannot be reached before the gas dissipates from the young planetary system. It should be noted, however, as was discussed earlier, that icy/rock low-mass planets could form by {\it DI} if there is 
 gaseous mass loss due to tidal stripping or photo-evaporation. The efficiency of forming planets like Uranus and Neptune in such a scenario is still unknown, and further investigation of this topic is required. \par

Modeling the interior structures of Uranus and Neptune is even more difficult.  In the first place, the high-Z component is a much larger fraction of the total mass and must therefore be treated more carefully.  In the second place, the thermal structure is not well known.  Neptune has a strong internal heat source, so an adiabatic interior is a reasonable assumption, but Uranus is in equilibrium with solar insulation (e.g., {\it Pearl and Conrath}, 1991).  In addition, thermal evolution models of Uranus give too long a cooling time for the planet ({\it Fortney et al.}, 2011).  It seems likely that some process like layered diffusion is inhibiting the heat flow in the planet, and that temperatures in the interior might be much higher than adiabatic.

In addition, the observational data for Uranus and Neptune have much larger uncertainties than for Jupiter and Saturn. The observed gravitational moment $J_2$, flattening $f$ and angular velocity $\omega$ are related to first order by 
\begin{equation}\label{rotation}
J_2-\frac{2f}{3}+\frac{q}{3}\approx 0
\end{equation}
where $q\equiv \omega^2a^3/GM$ is the ratio of centrifugal to gravitational forces.  The rotation periods determined by the Voyager flybys together with the measured values of $J_2$ and $f$ do not satisfy eq.\,\ref{rotation} ({\it Podolak and Helled}, 2012).  Indeed {\it Helled et al.} (2010) suggested that Uranus's rotation period is significantly shorter, and Neptune's significantly longer than the values determined by Voyager.

Uranus and Neptune have usually been treated using three-shell models, where the structure was divided into a core of rock, a shell of ``ice", and a H-He envelope in the solar ratio together with an admixture of high-Z material.~ More recent models have maintained the three-shell structure but have replaced the ice shell and envelope with an inner envelope containing H-He together with a large fraction of high-Z material and an outer envelope composed of H-He and a small mass fraction of high-Z material.

As is the case for Jupiter and Saturn, there is no compelling reason to limit the number of shells to three.  {\it Helled et al.} (2010) have assumed that 
$\rho(r)$ for these planets could be fit by a 6th-order polynomial, and have 
found the coefficients required to fit all of the observed parameters.  
The ``empirical" equation of state generated by these models could be 
interpreted as requiring a continuous increase in the H-He mass 
fraction with increasing radius.  This would be the equivalent of infinitely 
many shells.  Without a clearer understanding of the temperature structure 
inside these bodies it is difficult to generate more detailed models. 
In addition, {\it Helled et al.} (2010) have suggested that Uranus and 
Neptune could be rather "dry". It was shown that the gravity data can be fit 
as well with silicates as with water. This also raises the question whether Uranus 
and Neptune are truly icy planets, or alternatively, silicate-dominated planets, similar to Pluto ({\it Simonelli and Reynolds}, 1989). 

A puzzling problem to face modelers is the fact that although Uranus and Neptune have similar masses and radii, and are both found in the outer regions of the solar system, they appear to be quite different internally.  In addition to the differences in heat flow mentioned above, Uranus's radius is larger than Neptune's but it is smaller in mass. This means that Neptune is more dense than Uranus by 30\%.  This difference is likely to be the result of differences in the formation history of these bodies.  {\it Stevenson} (1986) and {\it Podolak and Helled} (2012) have discussed the possibility that giant impacts have significantly affected the internal structure of these planets, and have led to the observed dichotomy between them.  If this is so, then details of this formation history may eventually be extracted from interior models.  This exciting possibility is still not within our grasp, and should also be considered when modeling interiors of exoplanets with similar masses.

\noindent
\subsection{\textbf{Giant Exoplanets}}
\subsubsection{\textbf{Structure and Composition of Exoplanets}}
\subsubsection*{Basic Concepts}
There are two main ways that one might try to constrain the composition of a giant exoplanet.  Since gravity field or moment of inertia information about distant planets will be rarely available, if at all, observers must utilize more indirect methods.  One avenue is from spectroscopy.  Since our Solar System's giant planets are enhanced in heavy elements (at least in carbon, which can be observed as methane), spectra of exoplanet atmospheres could potentially also be used to determine if the H/He envelopes of such planets are enriched in heavy elements ({\it Chabrier et al.}, 2007; {\it Fortney et al.}, 2008).  In warm planets, H$_2$O, CO, CH$_4$, and NH$_3$ could all be detectable in the infrared.  However, high S/N observations will be needed across a wide wavelength range for strong constraints, either for hot Jupiters observed during transit or occultation, or young planets that are directly imaged.  There are no firm constraints on giant planet envelope metallicity at this time.

A simpler avenue, at least conceptually, is direct observations of a 
transiting planet's mass and radius, which yield its mean density.  Models 
for a solar composition giant planet at a given measured mass and age can be 
calculated.  If the observed planet has a smaller radius (has a higher 
mean density) than the model planet, this is strong evidence that the 
planet is enhanced in heavy elements compared to the solar composition model.
It is readily apparent that Jupiter and Saturn are enhanced in heavy elements compared to the Sun, from a measurement of their mass and radius alone, without knowledge of their gravity field.  Of course for an exoplanet, we will generally remain ignorant of whether these heavy elements are predominantly within a core or mixed throughout the H/He envelope.

\subsubsection*{Recent Work}
The use of mass and radius measurements of transiting planets as a structure probe for giant planets has been strongly hindered in practice. Most strongly irradiated planets (\teq$>1000$K) have radii larger than expected from standard models of giant planet cooling and contraction ({\it Bodenheimer et al.}, 2001; {\it Guillot and Showman}, 2002; {\it Baraffe et al.}, 2003; {\it Fortney et al.}, 2007; {\it Burrows et al.}, 2007).  
Some un-modeled physical process yields planetary radii (and hence, interior temperatures) that are inflated compared to the predictions of simple contraction models.  Since for any given planet the level of additional interior energy (or other process) that causes radius inflation is unknown, the amount of heavy elements within a planet (that would cause a smaller radius) is therefore unconstrained.  In cases where the planet is still smaller than a pure H/He object, a lower limit on its heavy element mass can be obtained.

Even with the uncertainty inherent in the radius inflation mechanism, progress has been made on constraining exoplanet composition with transiting planets.  With a sample size of only nine planets {\it Guillot et al.} (2006) were able to show that there is a strong correlation between stellar metallicity and the inferred mass of heavy elements within a planet.  The most iron-rich stars possessed the most heavy element rich planets.  However, {\it Guillot et al.} (2006) made the ad hoc assumption that the radius inflation mechanism (additional power in the convective interior) scaled as the stellar flux incident upon each planet.  With a somewhat larger sample size {\it Burrows et al.} (2007) came to a similar relation between stellar and planetary metal enrichment.  {\it Burrows et al.} (2007) made the assumption that ad hoc high atmospheric opacity slowed the contraction of all planets, leading to larger-than-expected radii at Gyr ages.

{\it Miller and Fortney} (2011) recognized that anomalously large radii disappear at incident fluxes less than $2\times10^8$ erg s$^{-1}$ cm$^{-2}$ (\teq$\sim1000$K).  By treating only these relatively cool planets, and assuming no additional internal energy source for them, the heavy element mass of each planet can be derived, as \emph{all} of these cooler planets are smaller than solar-composition H-He objects.  With their sample of 14 planets, they find that all need at least $\sim$10-15 \me\ of heavy elements. 
The enrichments of Saturn and Jupiter were found to fit in well within their exoplanetary ``cousin'' planets.  
When the heavy element mass is plotted vs.~metallicity, there is a clear correlation between the mass of heavy elements in a giant planet and the metallicity of the host star as already noted by {\it Guillot et al.} (2006) and {\it Burrows et al.} (2007).  However, the enrichment (i.e. $Z_{\mathrm{planet}}$/$Z_{\mathrm{star}}$) is not very sensitive to the metallicity of the star. The data suggest that it is the {\it planetary mass} that determines the relative enrichment and not the metallicity of the star. This tendency is consistent with the finding that the occurrence rate of low-mass exoplanets is not increasing with increasing stellar metallicity ({\it Buchhave et al.}, 2012). The sample implies that for low metallicity ([Fe/H]$<$0) the planets which exist are the ones with the large enrichments, i.e., more terrestrial planets. For higher metallicities ([Fe/H]$>$0), a large range of enrichments is possible. 
Clearly, a new area of modeling work will be the comparison of population 
synthesis models (see chapter by Benz et al.~) to observed giant planet \emph{composition} vs.~planetary mass, stellar mass, stellar metallicity, orbital separation, eccentricity, orbital alignment, and other parameters.  As the sample size of cooler transiting gas giants expands, such avenues of research are likely to bear fruit.

One important finding from the past decade is that it is clearly incorrect to uniformly assume that all giant planets possess $\sim$15 \me\ of heavy elements, an estimate that would work fairly well for Jupiter, Saturn, Uranus, and Neptune.  The detection of transits of the relatively small radius Saturn-mass planet HD 149026 by {\it Sato et al.} (2005) showed that even relatively low mass ``gas giants'' can possess 60-90 \me\ of heavy elements.  Relatively small radii for massive giant planets from $\sim$~2--10 \mj\ show that hundreds of earth masses of heavy elements must be incorporated into some planets ({\it Baraffe et al.}, 2008; {\it Miller and Fortney} 2011).  It seems likely that for massive giant planets with large enrichment in heavy elements, the bulk of the heavy element mass is in the H/He envelope, rather than in a core.  This is an area that should be investigated in future formation models.

\subsubsection*{Degeneracy in Composition}
While it is certainly essential to derive the heavy element mass fractions for giant planets, it would be even more interesting to derive additional details, like the ice/rock ratio.  This could in principle be easier for the lower mass (Neptune-like) giant planets where the bulk of their mass is in heavy elements, not H/He.  However, there is significant compositional degeneracy, as rock-H/He mixtures yield similar mass-radius relations as three-component models that include rock, ices, and H/He.  This is not a new phenomenon, and as discussed earlier, it has long been appreciated that Neptune and Uranus can be modeled with a wide array of mixtures of rock, ices, and H/He (e.g., {\it Podolak et al.} 1991).  Therefore there is no observational evidence that exoplanets in the ice giant mass range are actually composed mostly of fluid ices.  There is at least one possible way around this degeneracy that could yield some information on heavy element composition.  One could obtain correlations for a large number of planets for planetary heavy element enrichment as a function of stellar [O/H] and [Si/H], in addition to [Fe/H].

\section{\textbf{THE FUTURE}}
Observations of hundreds of exoplanets, as well as increasingly detailed information about the giant planets in our own Solar System have inspired modelers to explore planet formation scenarios in more detail.  To date we have explored the effects of such processes as migration, planetesimal capture, disk metallicity, disk magnetic fields etc.  It is perhaps not surprising that these processes often work in opposite directions. We are entering a new stage in our understanding of giant planets: one where we realize that in order to untangle the complex interplay of the relevant physical processes it is necessary to build models that self-consistently combine various physical processes.  In particular, we suggest the following specific investigations:
\begin{itemize}

\item Work toward a self-consistent model for the evolution of the growing
planet, the gaseous disk, and the planetesimal distribution in the {\it CA}
paradigm.
\vspace{-1.5mm}
\item Develop a better model for planetesimal properties: velocities,
inclinations, sizes, compositions, morphology, and the time dependence of these
properties.
\vspace{-4.5mm}
\item Track the deposition of heavy elements that are accreted in the envelope of a forming planet
and account for their EOS and opacity in {\it CA} models.
\vspace{-1.5mm}
\item Determine the final masses and identify the primary physical effects 
which determines the range of final masses in the {\it CA}, {\it DI}, and hybrid paradigms.
\vspace{-1.5mm}
\item Develop opacity calculations which account for the initial size distribution of interstellar grains, grain fragmentation, various possible morphologies, and grain composition for both {\it CA} and {\it DI}.
\vspace{-1.5mm}
\item Develop self-consistent models of the disk-planet interaction and 
co-evolution in the {\it DI} model. 
\vspace{-1.5mm}
\item Determine whether {\it in situ} formation of short-period massive and intermediate-mass planets
is plausible.
\vspace{-1.5mm}
\item Explore the earliest stages of disk evolution to search for
observational signatures of planet formation mechanisms. For example,
if {\it DI} is a viable formation pathway, then spiral
structure in very young disks should be detectable. ÊALMA and
future instruments may make this possible.
\vspace{-1.5mm}
\item Determine what physical factors lead to fragmentation, how
metallicity affects planet formation by {\it DI}, and the observational consequences for the {\it DI} model.
\vspace{-5.5mm}
\item Perform high-resolution 2D/3D simulations of clumps through their contraction
phases, including the effects of rotation, tidal perturbations, and
background pressures and temperatures. 
\vspace{-1.5mm}
\item Develop predictions of the two formation models that can be tested by future observations. 
\vspace{-2.5mm}
\item Perform 2D structure model calculations for the Solar System planets accounting for the static and dynamical contributions, improved equations of state of H, He, high-Z and their interactions, and possibly, magnetic fields.
\vspace{-2.5mm}
\item Include additional constraints for the planetary
internal structure of giant exoplanets such as the Love number/flattening and atmospheric
composition. 

\end{itemize}
This work is now beginning and it is hoped that by PPVII we will understand planetary systems better, both our own and those around other stars.

\section*{\textbf{Acknowledgments}}
{\small 
{\small 
A.C.B's support was provided under contract with the California
Institute of Technology funded by NASA via the Sagan Fellowship
Program. A.P.B's work was partially supported by the NASA Origins of Solar
Systems Program (NNX09AF62G). F.M. is supported by the ETH Zurich Post-doctoral Fellowship Program and by the Marie Curie Actions for People COFUND program.  
P. B. was supported in part by a grant from the NASA Origins program.   M.P. acknowledges support from ISF grant 1231/10.
Y.A. is supported by the European Research Council through grant 239605 and thanks the International Space Science Institute, in the framework of an ISSI Team.
}

\centerline\textbf{REFERENCES}
\parskip=0pt
{\small
\baselineskip=11pt

\refs Alexander R. D. et al.~(2008). 
{\em Mon. Not. R. Astron. Soc.}, {\em 389}, 1655--1664.

\refs Alibert Y. et al.~(2004). {\it Astron. Astrophys.}, {\it 417}, L25--L28. 

\refs Alibert Y. et al.~(2005a). {\it Astron. Astrophys.}, {\it 434}, 343--353. 

\refs Alibert Y. et al.~(2005b). {\it Astrophys. J.}, {\it 626}, L57--L60. 

\refs Alibert Y. et al.~(2011). {\it Astron. Astrophys.}, {\it 526}, A63.

\refs Anderson J. D. and Schubert G. (2007). {\it Science}, {\it 317}, 1384--1387. 
 
\refs Andrews S. M. et al.~(2011). {\it Astrophys. J.}, {\it 732}, 42.

\refs Andrews S. M. et al.~(2013). {\it Astrophys. J.}, in press, arXiv: 1305.5262 

\refs Ayliffe B. and Bate M. R. (2012). \textit{Mon. Not. R. Astron. Soc.}, \textit{427}, 2597--2612.

\refs {Baraffe} I. et al.~(2003). {\it Astron. Astrophys.}, {\it 402}, 701--712.

\refs {Baraffe} I. et al.~(2008). {\it Astron. Astrophys.}, {\it 482}, 315--332.

\refs Baruteau C. et al.~(2011). {\it Mon. Not. R. Astron. Soc.}, {\it 416}, 1971--1982.

\refs Batygin K. and Stevenson D. J. (2010).  \textit{Astrophys. J. }, \textit{714}, L238--L243.
  
\refs  Bodenheimer P. et al.~(1980) .\textit{Icarus}, \textit{41}, 293--308.

\refs Bodenheimer P. and Pollack J. B. (1986). \textit{Icarus}, \textit{67}, 391--408.

\refs  Bodenheimer P. et al.~(2000). 
\textit{Icarus}, \textit{143}, 2--14.     

\refs Bodenheimer P. et al.~(2001). \textit{Astrophys. J.}, \textit{548}, 466--472.

\refs Bodenheimer P. et al.~(2013). {\it Astrophys. J.}, {\it 770}, 120.

\refs Boley A. C. et al.~(2007). {\em Astrophys. J.}, {\em 665}, 1254--1267. 

\refs Boley A. C. et al.~(2010). {\it Icarus}, {\it 207}, 509--516.

\refs Boley A. C. and Durisen R. H. (2010). {\it Astrophys. J.}, {\it 724}, 618--639. 

\refs Boley A. C. et al.~(2011). {\it Astrophys. J.}, {\it 735}, 30. 

\refs Boss A. P. (2002). {\it Astrophys. J.}, {\it 567}, L149--L153. 

\refs Boss A. P. et al.~(2002). {\it Icarus}, {\it 156}, 291--295. 

\refs Boss A. P. (2006a). {\it Astrophys. J.}, {\it 641}, 1148--1161.

\refs Boss A. P. (2006b). {\it Astrophys. J.}, {\it 644}, L79--L82.

\refs Boss A. P. (2009). {\it Astrophys. J.}, {\it 694}, 107--114.

\refs Boss A. P. (2011). {\it Astrophys. J.}, {\it 731}, 74--87.

\refs Boss A. P. (2012). {\it Mon. Not. R. Astron. Soc.}, {\it 419}, 1930--1936.

\refs Bromley B. C. and Kenyon S. J. (2011). {\it Astrophys. J.}, {\it 731}, 101--119.

\refs Buchhave L.~A. et al.~(2012). {\it Nature}, {\it 486}, 375--377.

\refs {Burrows} A. et al.~(2007). {\it Astrophys. J.}, {\it 661}, 502--514.

\refs Cai K. et al.~(2006). {\it Astrophys. J.}, {\it 636}, L149--L152. 

\refs Cai K. et al.~(2008). {\it Astrophys. J.}, {\it 673}, 1138--1153. 

\refs Cha S. H. and Nayakshin S. (2011).  {\em Mon. Not. R. Astron. Soc.} {\em 415}, 3319--3334.

\refs Chabrier G. et al.~(1992). {\it Astrophys. J.}, {\it 391}, 817--826.

\refs Chabrier G. et al.~(2000). 
\textit{Astrophys. J.}, \textit{542}, 464--472.

\refs {Chabrier} G. et al.~(2007). In {\it Protostars and Planets V}, ed. B.~{Reipurth},
  D.~{Jewitt} and K.~{Keil}, (Tucson: Univ. Arizona Press), 623--638.
  
\refs Clarke C. J. et al.~(2007).  {\em Mon. Not. R. Astron. Soc.}, {\em 381}, 1543--1547.

\refs Clarke C. J. (2009). {\it Mon. Not. R. Astron. Soc.}, {\it 396}, 1066. 

\refs Clarke C. J. and Lodato G. (2009). {\it Mon. Not. R. Astron. Soc.}, {\it 398}, L6--L10. 

\refs Cossins, P. et al.~(2010). {\it Mon. Not. R. Astron. Soc.}, {\it 401}, 2587--2598. 

\refs Crida A. et al.~(2006). {\it Icarus}, {\it 181}, 587--604.

\refs  D'Angelo G. et al.~(2011). In \textit{{\bf exoplanets}}, ed. S. Seager (Tucson: Univ. Arizona Press).

\refs DeCampli W. M. and Cameron A. G. W. (1979). {\it Icarus}, {\it 38}, 367--391. 

\refs Demarcus W. C. (1958). {\it Astron. J.}, {\it 63}, 2--28.

\refs Durisen R. H. et al.~(1986). {\it Astrophys. J.}, {\it 305}, 281--308. 

\refs Durisen R. H. et al.~(2007). In {\em 
Protostars and Planets V}, ed. B. Reipurth, D. Jewitt and K. Keil (Tucson:
Univ. Arizona Press), 607--621. 

\refs  Fischer D. A. and Valenti J. (2005). \textit{Astrophys. J.}, \textit{622}, 1102--1117.

\refs Forgan D. and Rice K. (2011). {\it Mon. Not. R. Astron. Soc.}, {\it 417}, 1928--1937. 

\refs Fortier A. et al.~(2013). {\it Astron. Astrophys.}, {\it 549}, A44.

\refs {Fortney} J.~J. et al.~(2007). {\it Astrophys. J.}, {\it 659}, 1661--1672.

\refs {Fortney} J.~J. et al.~(2008). {\it Astrophys. J.}, {\it 683}, 1104--1116.

\refs Fortney J. J. and Nettelmann N. (2010). {\it Sp. Sci. Rev.}, {\it 152}, 423--447.
  
\refs {Fortney} J.~J. et al.~(2011). {\it Astrophys. J.}, {\it 729}, 32--46.

\refs {{Galvagni} M. et al.~(2012). {\it Mon. Not. R. Astron. Soc.}, {\it 427}, 1725--1740. 

\refs Gammie C. F. (2001). {\it Astrophys. J.}, {\it 553}, 174--183. 

\refs Gautier D. et al.~ (2006). 
In {\it 36th COSPAR Scientific Assembly}, abstract 867.

\refs Gonzalez G. (1997).  {\it Mon. Not. R. Astron. Soc.}, {\it 285}, 403.

\refs Greenzweig Y. and Lissauer J. (1992). {\it Icarus}, \textit{100}, 440--463.

\refs Guillot T. et al.~(1994). {\it Icarus}, {\it 112}, 337--353.

\refs Guillot T. (1999). {\it Planet. Space Sci.}, {\it 47}, 1183--1200.

\refs {Guillot} T. and {Showman} A.~P. (2002). {\it Astron. Astrophys.}, {\it 385}, 156--165.

\refs {Guillot} T. et al.~(2006). {\it Astron. Astrophys.}, {\it 453}, L21--L24.

\refs Hartmann L. and Kenyon S. J. (1996). {\it Annu. Rev. Astron. Astrophys.}, {\it 34}, 207--240.

\refs Hellary P. and Nelson R. P. (2012). {\it Mon. Not. R. Astron. Soc.}, {\it 419}, 2737. 

\refs Helled R. et al.~(2006). {\it Icarus}, {\it 185}, 64--71.

\refs Helled R. et al.~(2008). {\it Icarus}, {\it 195}, 863--870.

\refs Helled R. and Schubert G. (2008). {\it Icarus}, {\it 198}, 156--162.

\refs Helled R. and Schubert G. (2009). {\it Astrophys. J.}, {\it 697}, 1256--1262.

\refs Helled R. and Bodenheimer P. (2010). {\it Icarus}, {\it 207}, 503--508. 

\refs Helled R. et al.~(2010). {\it Icarus}, {\it 210}, 446--454.

\refs Helled R. and Bodenheimer P. (2011). {\it Icarus}, {\it 211}, 939--947. 

\refs Helled R. et al.~(2011). {\it Astrophy. J.}, {\it 726}, 1--15.

\refs Hillenbrand L. (2008). \textit{Phys. Script.}, \textit{130}, 014024.   

\refs Hori Y. and Ikoma M. (2011). {\it Mon. Not. R. Astron. Soc.}, {\it 416}, 1419--1429. 
          
\refs Hubbard W. B. (1968). {\it Astrophys. J.}, {\it 152}, 745--754.

\refs  Iaroslavitz E. and Podolak M. (2007). \textit{Icarus}, \textit{187}, 600--610.
         
\refs Ida S. and Lin D. N. C. (2004). \textit{Astrophys. J.}, \textit{616}, 567--572.   
         
\refs  Ida S. and Lin D. N. C. (2005). \textit{Astrophys. J.}, \textit{626}, 1045--1060.   

\refs  Inaba S. et al.~(2003). \textit{Icarus}, \textit{166}, 46--62.
                   
\refs Johnson J. A. et al.~(2010). \textit{Publ. Astron. Soc. Pacific}, \textit{122}, 905--915.

\refs Kennedy G. M. and Kenyon S. J. (2009). \textit{Astrophys. J.}, \textit{695}, 1210--1226.
                   
\refs Kley W. and Dirksen G. (2006). {\it Astron. Astrophys.}, {\it 447}, 369--377. 

\refs Kley W. et al.~(2009). {\it Astron. Astrophys.}, {\it 506}, 971--987.

\refs Kobayashi H. et al.~(2010). {\it Icarus}, {\it 209},  836--847.
 
\refs Kratter K. M. et al.~(2010a). {\em Astrophys. J.}, {\em 708}, 1585--1597.

\refs Kratter K. M. et al.~(2010b).  {\em Astrophys. J.}, {\em 710}, 1375--1386. 
 
\refs Laughlin G. et al.~(2004). \textit{Astrophys. J.}, \textit{612}, L73--L76.   
 
\refs Leconte J. and Chabrier G. (2012). {\it Astron. Astrophys.}, {\it 540}, A20.
 
\refs  Lin D. N. C. and Papaloizou J. C. B. (1979).  \textit{Mon. Not. R. Astron. Soc.}, \textit{186}, 799--812. 

 \refs  Lin D. N. C. et al.~(1996). \textit{Nature}, \textit{380}, 606--607.

\refs  Lissauer J. J. et al.~ (2009). \textit{Icarus},\textit{199}, 338--350.

\refs Lodato G. and Clarke C. J. (2011). {\it Mon. Not. R. Astron. Soc.}, {\it 413}, 2735--2740. 

\refs Lubow S.~H. and D'Angelo G. (2006). {\em Astrophys. J.}, {\it 641}, 526--533.

\refs Maldonado J. et al.~(2013). {\it Astron. Astrophys.}, {\it 544}, A84. 

\refs Mamajek E. E. (2009).  In 
{\it {\bf exoplanets} and Disks: their Formation and Diversity:  AIP Conference 
Proceedings}, {\it 1158},  3--10.

\refs Marley M. et al.~(2007). {\em Astrophys. J}, {\it 655}, 541--549.

\refs Marois C. et al.~ (2008). {\it Science}, {\it 322}, 1348--1352.

\refs Mayer L. et al.~(2004).  {\em Astrophys. J}, {\it 609}, 1045. 

\refs Mayer L. et al.~(2007). {\em Astrophys. J.}, {\em 661}, L77--L80. 

\refs Mayer L. et al.~(2007). {\it Science}, {\it 316}, 1874--1877. 

\refs Mayer L. and Gawryszczak A J. (2008). In: {\it Extreme Solar Systems}, ed. D. Fischer et al.~(San Francisco: Astron.~Soc.~Pac.~), 243. 

\refs Mayor M. et al.~(2011). arXiv: 1109.2497.

\refs Meru F. and Bate M. R. (2010). {\em Mon. Not. R. Astron. Soc.}, {\em 406}, 2279--2288.

\refs Meru F. and Bate M. R. (2011a).  {\em Mon. Not. R. Astron. Soc.}, {\it 411}, L1--L5.

\refs Meru F. and Bate M. R. (2011b).  {\em Mon. Not. R. Astron. Soc.}, {\it 410}, 559--572.

\refs Meru F. and Bate M. R. (2012). {\em Mon. Not. R. Astron. Soc.}, {\em 427}, 2022--2046.

\refs Michael S. and Durisen R. H. (2010). {\em Mon. Not. R. Astron. Soc.}, {\em 406}, 279--289.

\refs Michael S. et al.~(2011). {\it Astrophys. J.}, {\it 737}, 42--48. 

\refs Militzer B. et al.~(2008). {\it Astrophys. J.}, {\it 688}, L45--L48.

\refs {Miller} N. and {Fortney} J.~J. (2011). {\it Astrophys. J.}, {\it 736}, L29--L34.

\refs  Mizuno H. (1980). \textit{Prog. Theor. Phys.}, \textit{64}, 544--557.          

\refs Mordasini C., Alibert Y. and Benz W. (2009a). {\it Astron. Astrophys.}, {\it 501}, 1139--1160. 

\refs Mordasini C. et al.~(2009b). {\it Astron. Astrophys.}, {\it 501}, 1161--1184. 

\refs Mordasini C. et al.~(2012a). {\it Astron. Astrophys.}, {\it 541} ,A97. 

\refs Mordasini C. et al.~(2012b). {\it Astron. Astrophys.}, {\it 547}, A111. 

\refs Mordasini C. (2013). {\it Astron. Astrophys.}, in press. 
         
\refs Mortier A. et al.~(2013). 
{\it Astron. Astrophys.}, {\it 551}, A112.          
         
\refs Movshovitz N. et al.~(2010). \textit{Icarus},
 \textit{209}, 616--624.  

\refs  M{\"u}ller T. W. A. et al.~(2012). {\it Astron. Astrophys.}, {\it 541}, A123. 

\refs Nayakshin S. (2010). {\it Mon. Not. R. Astron. Soc.}, {\it 408}, 2381--2396.

\refs Nayakshin S. (2011). {\it Mon. Not. R. Astron. Soc.}, {\it 416}, 2974--2980.

\refs Nayakshin S. and Cha S. H. (2012). {\it Mon. Not. R. Astron. Soc.}, {\it 423}, 2104--2119.

\refs Nayakshin S. and Lodato G. (2012). {\it Mon. Not. R. Astron. Soc.}, {\it 426}, 70--90.

\refs Nelson A. F. et al.~(1998). {\it Astrophys. J.}, {\it 502}, 342--371. 

\refs Nelson A. F. (2006). {\em Mon. Not. R. Astron. Soc.}, {\em 373}, 1039--1070.

\refs Nettelmann N. et al.~(2008). {\it Astrophys. J.}, {\it 683}, 1217--1228.

\refs Nettelmann N. et al.~(2012). {\it Astrophys. J.}, {\it 750}, 52.

\refs Paardekooper S. et al.~(2011).  {\em Mon. Not. R. Astron. Soc.}, {\em 416}, L65--L69.

\refs Pearl J. C. and Conrath B. J. (1991). {\it J. Geophys. Res.}, {\it 96}, 18921--18930.

\refs Pickett M. K. and Durisen R. H. (2007), {\em Astrophys. J.}, {\em 654}, L155--L158.

\refs Podolak M. and Cameron A. G. W. (1974). {\it Icarus}, {\it 22}, 123--148. 

\refs  Podolak M. et al.~(1988). \textit{Icarus}, \textit{73}, 163--179.

\refs Podolak M. and Hubbard W. B. (1998).  In: {\it Solar System Ices}, ed. M. F. B. Schmitt, and C. de Bergh (Dordrecht: Kluwer), 735--748.

\refs {Podolak} M. et al.~(1991). In: {\it Uranus}, ed. J.T. Bergstralh, E. D. Miner and M. S. Matthews (Tucson: Univ. Arizona Press),  29--61.

\refs Podolak M. (2003).  \textit{Icarus}, {\it 165}, 428--437.

\refs Podolak M. et al.~(2011). {\it Astrophys. J.}, {\it 734}, 56. 

\refs Podolak M. and Helled R. (2012). {\it Astrophys. J.}, {\it 759}, L32.

\refs Pollack J. B. et al.~(1996). \textit{Icarus}, \textit{124}, 62--85.          

\refs Rafikov R. R. (2007). {\em Astrophys. J.}, {\em 662}, 642--650.

\refs Rafikov R. R.  (2011). {\it Astrophys. J.}, {\it 727}, 86.

\refs Rice W. K. M. et al.~(2004). {\it Mon. Not. R. Astron. Soc.}, {\it 355}, 543--552. 

\refs Rice W. K. et al.~(2005). {\it Mon. Not. R. Astron. Soc.}, {\it 364}, L56--L60. 

\refs Rice W. K. M. et al.~(2006). {\it Mon. Not. R. Astron. Soc.}, {\it 372}, L9-L13. 

\refs Redmer R. et al.~(2011). {\it Icarus}, {\it 211}, 798--803.

\refs Robinson S. E. et al.~(2006).  \textit{Astrophys. J.}, \textit{643}, 484--500.      

\refs Rogers P. D. and Wadsley J. (2012). {\it Mon. Not. R. Astron. Soc.}, {\it 423}, 1896--1908.        
        
\refs Safronov V. S. (1969). \textit{Evolution of the Protoplanetary Cloud and Formation of the Earth and Planets} (Moscow: Nauka), in
         Russian. 
         
\refs Santos N. C. et al.~(2004).  
{\em Astron. Astrophys.}, {\em 415}, 1153--1166.

\refs Santos N. C. et al.~(2011).   
{\em Astron. Astrophys.}, {\em 526}, A112.
         
\refs Sato B. et al.~(2005). {\it Astrophys. J.}, {\it 633}, 465--473.

\refs Saumon D. et al.~(1995). \textit{Astrophys. J. Suppl.}, \textit{99}, 713--741.  

\refs Saumon D. and Guillot T. (2004). \textit{Astrophys. J.}, \textit{609}, 1170--1180.

\refs Simonelli D. P. and Reynolds R. T. (1989). {\em Space Sci. Rev.}, {\it 16}, 1209--1212.  

\refs Stamatellos D. and Whitworth A. P. (2008). {\em Astron. Astrophys.}, {\em 480}, 879--887.

\refs Stanley S. and Bloxham J. (2006). {\em Icarus}, {\it 184}, 556--572.

\refs Stevenson D. J. (1975). {\it Phys. Rev. Condens. Matter B}, {\it 12}, 3999--4007. 

\refs Stevenson D. J. (1982). \textit{Planetary and Space Sci.} \textit{30}, 755--764.  

\refs Toomre A. (1964). {\it Astrophys. J.}, {\it 139}, 1217--1238.

\refs Tsiganis K. et al.~(2005). 
\textit{Nature}, \textit{435}, 459--461.

\refs Uribe A. L. et al.~(2013).  {\it Astrophys. J.}, 
{\it 769}, 97.

\refs Vazan A. and Helled R. (2012). {\it Astrophys. J.}, {\it 756}, 90.

\refs Vorobyov E. I. and Basu S. (2006). {\it Astrophys. J.}, {\it 650}, 956--969.

\refs Ward W. R. (1997). \textit{Astrophys. J.}, \textit{482}, L211--L214.

\refs  Weidenschilling S. J. (1977). {\it Astrophys. Space Sci.}, {\it 51}, 153--158.   
          
\refs  Weidenschilling S. J. (2008). \textit{Phys. Script.} \textit{130}, 014021.

\refs Weidenschilling S. J. (2011). {\it Icarus}, {\it 214}, 671--684.

\refs Wilson H. F. and Militzer B. (2010). {\it Phys. Rev. Lett.}, {\it 104}, 121101.

\refs Wuchterl G. et al.~(2000). In: {\it 
Protostars and Planets IV}, eds V. Mannings, A. P. Boss, S. S. Russell 
(Tucson: Univ. Arizona Press), 1081--1109. 

\refs Yasui C. et al.~(2009). {\it Astrophys. J.}, {\it 705}, 54--63. 

\refs Zhu Z et al.~(2012).  {\it Astrophys. J.}, {\it 758}, L42. 
  
\end{document}